\documentclass[twocolumn]{autart}    

\usepackage{amsmath}
\usepackage{amsfonts}
\usepackage{amssymb}
\usepackage{graphicx}
\usepackage{upgreek}
\usepackage{nicefrac}
\newcommand{\abs}[1]{\left| #1 \right|}
\newcommand{\norm}[1]{\left|\left| #1 \right|\right|}

\newcommand{\sign}{\textup{sign}}

\begin{document}

\begin{frontmatter}

\title{Passivity-based control of underactuated mechanical systems with Coulomb friction: \\Application to earthquake prevention\thanksref{footnoteinfo}} 

\thanks[footnoteinfo]{This paper was not presented at any IFAC meeting. Corresponding author Ioannis Stefanou.}

\author[Gem]{Diego Gutierrez-Oribio}\ead{diego.gutierrez-oribio@ec-nantes.fr},    
\author[Gem]{Ioannis Stefanou}\ead{ioannis.stefanou@ec-nantes.fr},               
\author[Ls2n]{Franck Plestan}\ead{franck.plestan@ec-nantes.fr}  

\address[Gem]{Nantes Universit\'e, \'Ecole Centrale Nantes, CNRS, GeM, UMR 6183, F-44000 Nantes, France}  
\address[Ls2n]{Nantes Universit\'e, \'Ecole Centrale Nantes, CNRS, LS2N, UMR 6004, F-44000 Nantes, France}             
          

\begin{abstract}                          
Passivity property gives a sense of energy balance. The classical definitions and theorems of passivity in dynamical systems require time invariance and locally Lipschitz functions. However, these conditions are not met in many systems. A characteristic example is nonautonomous and discontinuous systems due to presence of Coulomb friction. This paper presents an extended result for the negative feedback connection of two passive nonautonomous systems with set-valued right-hand side based on an invariance-like principle. Such extension is the base of a structural passivity-based control synthesis for underactuated mechanical systems with Coulomb friction. The first step consists in designing the control able to restore the passivity in the considered friction law, achieving stabilization of the system trajectories to a domain with zero velocities. Then, an integral action is included to improve the latter result and perform a tracking over a constant reference (regulation). At last, the control is designed considering dynamics in the actuation. These control objectives are obtained using fewer control inputs than degrees of freedom, as a result of the underactuated nature of the plant. The presented control strategy is implemented in an earthquake prevention scenario, where a mature seismogenic fault represents the considered frictional underactuated mechanical system. Simulations are performed to show how the seismic energy can be slowly dissipated by tracking a slow reference, thanks to fluid injection far from the fault, accounting also for the slow dynamics of the fluid's diffusion.
\end{abstract}

\end{frontmatter}

\section{Introduction}
\label{sec:Introduction}

Passivity is an important property in dynamical systems because it gives a sense on the system energy balance \cite{b:9740597,b:Khalil2002}. Roughly speaking, a system is said to be passive if it cannot produce energy on its own, and can only dissipate the energy that is stored in it at any time. Friction is a dissipative mechanism that is ubiquitous in mechanical systems \cite{b:Armstromg-Dupont-Canudas-1994,b:Olsson-Astrom-Canudas-Gafvert-Lischinsky-1998}. 
Although friction may be a desirable property (as in brakes application), it can also lead to limit cycles, undesired stick-slip motion and instabilities. This last phenomenon can be explained qualitatively due to the competition of stored elastic energy and its dissipation via friction. If this stored energy can not be balanced by the frictional dissipation, then, an instability will be triggered. This is the case when the frictional force decreases with slip or slip-rate and can be explained through the loss of passivity of the system. The prevention of such instabilities is the main objective in this work.


Due to its energy dissipation nature, friction has been compensated in mechanical systems using passivity-based controllers. The passivity-based control term was introduced in \cite{b:ORTEGA1989877} and it has an important role in the control theory with applications to electric motors, power electronics, chemical processes and mechanical systems (see \cite{b:9740597,b:10.1080/00207178908953515,b:ORTEGA1989877,b:Spong-Vidyasagar}). For the case of totally actuated mechanical systems, one can mention \cite{b:Canudas-Kelly-2007} where a LuGre (dynamic) model of friction is compensated with an observer, or \cite{b:doi.org/10.1002/asjc.2718}, where the stabilization of a system with Coulomb friction is analysed using sliding-modes. For the case of a system having less control inputs than degrees of freedom (underactuated system), the Interconnection and Damping Assignment Passivity-based Control (IDA-PBC) presented in \cite{b:Ortega_TAC2002} was used in \cite{b:doi.org/10.1002/rnc.1622}, \cite{b:10.1177/1077546311408469} and \cite{b:Franco-2021} (with an adaptive IDA-PBC) to stabilize systems with dynamic, but not set-valued, frictional models. Furthermore, IDA-PBC requires the solution of partial differential equations (PDEs) in the control, which is cumbersome and in some cases a solution might not exist.

Despite the attractiveness of passivity concepts, the classical passivity theorems (\textit{e.g.}, \cite[Chapter 6]{b:Khalil2002}) do not include set-valued frictional systems (like the Coulomb friction), which is the focus of this work. Furthermore, the classical theorems do not include nonautonomous systems either. There exist some works dealing with the feedback interconnection of multivalued systems using convex analysis (see, \textit{e.g.}, \cite{b:Adly-Le-2014,b:Brogliato-2004,b:10.1137/18M1234795} and a very recent monograph \cite{b:Brogliato-2022}), yet they do not take into account nonautonomous systems. For this purpose, in this work we extend the classical theorem of passivity related to the negative feedback connection between two passive systems, in such a way to cover the general class of underactuated frictional, nonautonomous systems with set-valued right-hand side (RHS). This is accomplished by using an invariance-like principle \cite{b:Finogenko-2016,b:Kamalapurkar-Rosenfeld-Parikh-Teel-Dixon-2019}. 


Based on this theoretical result, a passivity-based control design is considered to restore the passivity property of a nonautonomous with set-valued RHS underactuated mechanical system. First, stabilization to a domain of zero velocities is obtained, recovering the passivity property by properly designing the underactuated control input. Then, a regulation result over constant references is obtained by augmenting the system with integral action. Finally, the actuator dynamics is considered and the control is designed to preserve the regulation result. The designed underactuated control is implemented in an earthquake prevention scenario of a seismic fault. This is an important and challenging example of a frictional underactuated system, where the designed control has to be able to dissipate the stored energy slowly, controlling the fast dynamics of an earthquake through a slow diffusion process. Simulations are presented to show how the passivity-based control is able to follow a slow reference dissipating slowly the stored energy, avoiding in this manner, an earthquake-like behaviour.

The outline of this work is as follows. The notation and useful definitions and existing theorems are presented in Section \ref{sec:Pre}. The passivity extension for the negative feedback connection between two nonautonomous discontinuous systems, the main theorem of this paper, is presented in Section \ref{sec:theorem}. The frictional underactuated mechanical system description, the link between passivity and the considered friction law and the control objectives are given in Section \ref{sec:Problem}. The structured design of the passivity-based control is detailed in Section \ref{sec:Control}. The presentation of the fault model and the numerical simulations are shown in Section \ref{sec:SimExp}. Finally, some concluding remarks are discussed in Section \ref{sec:Conclusions}. 

\section{Preliminaries}
\label{sec:Pre}


Consider the $n$-dimensional space $\Re^n$ with the Euclidean norm $\norm{\cdot}$. Elements of $\Re^n$ are interpreted as column vectors and $(\cdot)^T$ denotes the vector transpose operator. The identity matrix of dimension $n$ is denoted by $I_n$ or simply $I$, if the size can be trivially determined by the context. Let $v \in \Re^n$, be the function $\sign{(\cdot)}: \Re^n \rightarrow \Re^{n \times n}$, defined as $\sign{(v)}= \textup{diag}[\sign{(v_1)}, ..., \sign{(v_n)}]$, with $\mathrm{sign}(v_i)=\left\{\begin{array}{cc}
1 & v_i>0 \\ 
\left[-1,1\right] & v_i=0 \\ 
-1 & v_i<0
\end{array}   \right.$, \\ for all $i=1,...,n$ and the function $\abs{\cdot}: \Re^n \rightarrow \Re^n$ is defined as $\abs{v}=[\abs{v_1},...,\abs{v_n}]^T$.

Consider the state model given by
\begin{equation}
\begin{split}
  \dot{x} = f_1(x,u),\quad
  y = h(x,u),
\end{split}
\label{eq:sys}
\end{equation}
where $f_1:\Re^n \times \Re^p \rightarrow \Re^n$ is locally Lipschitz, $h:\Re^n \times \Re^p \rightarrow \Re^p$ is continuous, $f_1(0,0)=0$ and $h(0,0)=0$.

\begin{defn} \cite{b:100932,b:1101352,b:HILL1977377},\cite[Chapter 6]{b:Khalil2002}\cite{b:Willems-1972} 
System \eqref{eq:sys} is said to be passive if there exists a continuously differentiable positive semidefinite function $V(x)$ (called the storage function) such that the passivity map (\textit{i.e.} $u^Ty$) fulfils $u^Ty \geq \dot{V}(x)=\frac{\partial V}{\partial x}f_1(x,u)$ $\forall$ $(x,u)\in \Re^n \times \Re^p$.
\label{def:passivity}
\end{defn}

\begin{figure}[ht!]
  \centering 
  \includegraphics[width=4.5cm,height=2.5cm]{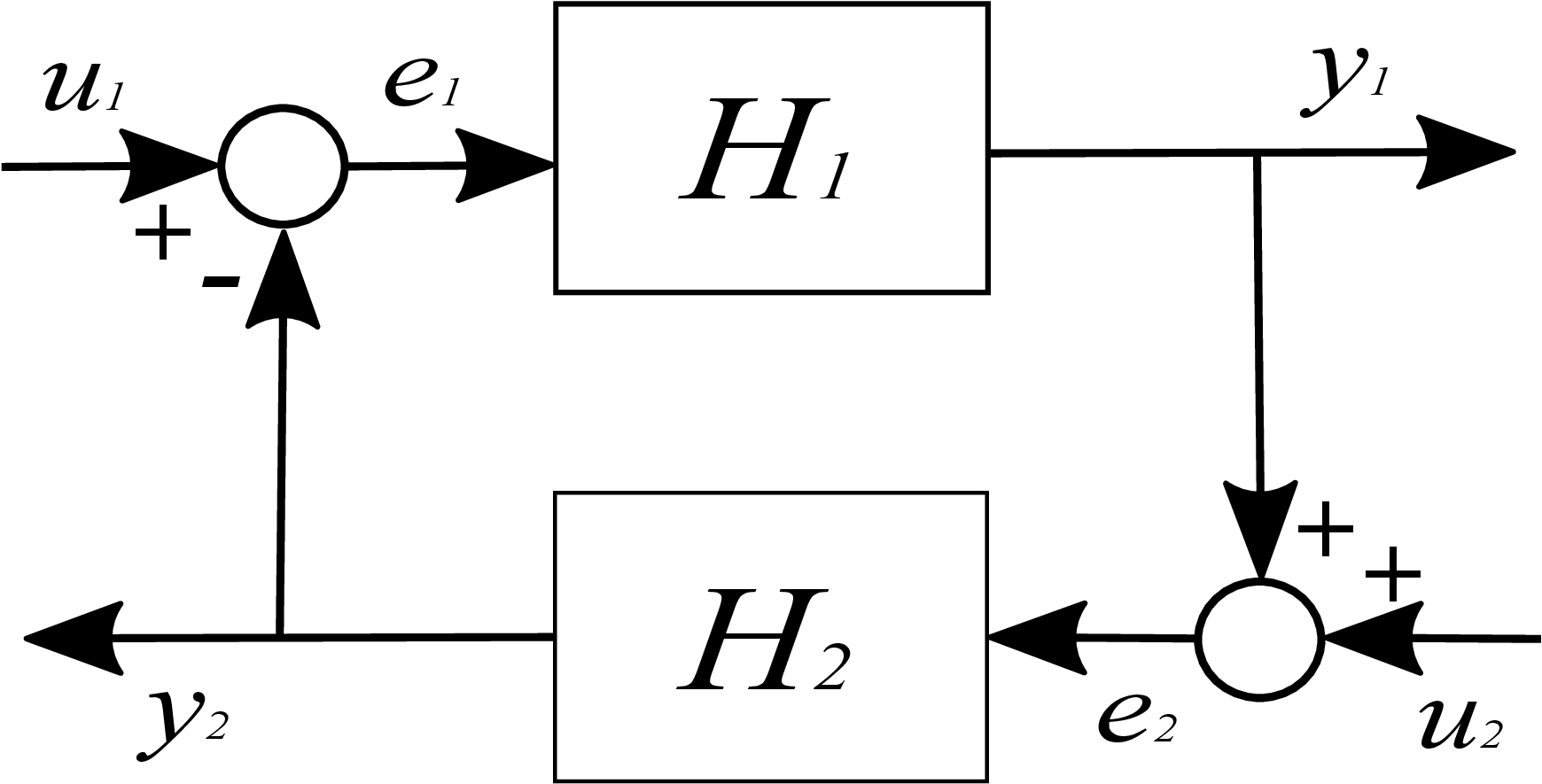}
  \caption{Negative feedback connection.}
  \label{fig:Feedback}
\end{figure}

An important passivity theorem concerns the negative feedback connection between two passive systems, $H_1$ and $H_2$ (Fig. \ref{fig:Feedback}). The systems $H_1$ and $H_2$ can be either time-invariant dynamical systems or (possibly time-variant) memoryless functions.

\begin{thm}\cite[Chapter 6]{b:Khalil2002}
If the system $H_1$ is passive with input $e_1$ and output $y_1$, and the system $H_2$ is passive with input $e_2$ and output $y_2$, then the negative feedback connection of $H_1$ and $H_2$ is passive with input $u=[u_1,u_2]^T$ and output $y=[y_1,y_2]^T$. 
\label{th:feedback}
\end{thm}

Let
\begin{equation}
\begin{split}
  \dot{x} &= f_2(x,t),\\
\end{split}
\label{eq:syst}
\end{equation}
where $f_2:\Re^{n} \times \Re_{\geq 0} \rightarrow \Re^n$  is piecewise continuous\footnote{A function is said to be piecewise continuous in a domain $G$ if it is continuous in $G$ up to a set of measure zero defined by points of discontinuity of the function.} in a domain $G \subset \Re^{n} \times \Re_{\geq 0}$. The above system is nonautonomous and has set-valued RHS (see \cite{b:Orlov-2009} for more details on discontinuous systems and \cite{b:10.1080/00207178708933801,b:4048815} for control robot manipulators with discontinuous RHS for time invariant and nonautonomous systems, respectively). In the following, the solutions of discontinuous systems like \eqref{eq:syst} are understood in the Filippov's sense \cite{b:filippov}.

\begin{thm}(\textbf{Invariance-like principle})\cite{b:Finogenko-2016,b:Kamalapurkar-Rosenfeld-Parikh-Teel-Dixon-2019}
Let $D \subset \Re^n$ be a domain containing $x=0$. Suppose there exists a constant $M$ such that $\norm{f_2(x,t)} \leq M$, for almost all $(x,t)\in D \times \Re$. Let $V:D \times \Re_{\geq 0} \rightarrow \Re$ be a locally Lipschitz-continuous positive definite function such that
\begin{equation*}
\begin{split}
  W_1(x) \leq V(x,t) \leq W_2(x), \\
  \dot{V}(x,t)=\frac{\partial V}{\partial t}+ \frac{\partial V}{\partial x} f_2(x,t) \leq -W(x),
\end{split}
\end{equation*}
for all $t \geq 0$ and for all $x \in D$, where $W_1(x)>0$, $W_2(x)>0$ and $W(x) \geq0 $ are continuous functions on $D$. Choose $r>0$ such that $B_r = \{\ x \in \Re^n \mid \norm{x} \leq r \}\ \subset D$ and let $\rho < \textup{min}_{\norm{x}=r}W_1(x)$. Then, every bounded Filippov solutions of system \eqref{eq:syst}, such that $x(t_0) \in \{\ x \in B_r \mid W_2(x)\leq \rho \}\ $ are bounded and satisfy $W(x(t))\rightarrow 0$ as $t \rightarrow \infty$. Consequently, $x(t)$ approaches $E= \{\ x \in D \mid W(x)=0 \}\ $ as $t \rightarrow \infty$. Moreover, if all assumptions hold globally and $W_1(x)$ is radially unbounded, the statement is true for all $x(t_0) \in \Re^n$.
\label{th:invariance}
\end{thm}

\section{Passivity Extension for Nonautonomous Discontinuous Systems}
\label{sec:theorem}

The classical definition of passivity in Definition \ref{def:passivity} does not consider directly systems in the form of \eqref{eq:syst} due to the time dependency and its set-valued RHS. For this purpose, we generalize Theorem \ref{th:feedback} for systems $H_1$ and $H_2$ that can be either time-variant dynamical systems with discontinuous RHS or time-variant discontinuous memoryless functions. Such theorem is the central result of this work and it will be used on each control design step.

\begin{thm}
Assume each element of the feedback interconnection of Fig. \ref{fig:Feedback} is passive and satisfies
\begin{equation*}
  e_i^T y_i \geq \dot{V}_i + \varphi_i(x), \quad \varphi_i: \Re^n \rightarrow \Re_{\geq 0}, \quad i =1,2.
\end{equation*}
Let $D \subset \Re^n$ be a domain containing $x=0$ and consider the locally Lipschitz-continuous positive storage function $V(x,t)=V_1(x,t)+V_2(x,t)$, $V:D \times \Re_{\geq 0} \rightarrow \Re$ such that $W_1(x) \leq V(x,t) \leq W_2(x)$, for all $t \geq 0$ and for all $x \in D$, where $W_1(x)>0$ and $W_2(x)>0$ are continuous on $D$. Choose $r>0$ such that $B_r = \{\ x \in \Re^n \mid \norm{x} \leq r \}\ \subset D$ and let $\rho < \textup{min}_{\norm{x}=r}W_1(x)$. Then, every bounded Filippov solutions of the closed-loop system shown in Fig. \ref{fig:Feedback} with $u_1=u_2=0$, \textit{i.e.} a system of the form \eqref{eq:syst}, such that $x(t_0) \in \{\ x \in B_r \mid W_2(x)\leq \rho \}\ $ are bounded and satisfy $W(x(t))\rightarrow 0$ as $t \rightarrow \infty$, with $W(x)=\varphi_1(x)+\varphi_2(x)$. Consequently, $x(t)$ approaches $E= \{\ x \in D \mid W(x)=0 \}\ $ as $t \rightarrow \infty$. Moreover, if all assumptions hold globally and $W_1(x)$ is radially unbounded, then the statement is true for all $x(t_0) \in \Re^n$.
\label{th:passive}
\end{thm}
\begin{pf}
Taking the function $V(x,t)=V_1(x,t)+V_2(x,t)$ as storage function of the closed-loop system, its derivative w.r.t. time is written as 
\begin{equation*}
\begin{split}
  \dot{V} &\leq -\varphi_1(x) - \varphi_2(x) + e_1^T y_1 + e_2^T y_2 \\
   &\leq -W(x) + (u_1-y_2)^T y_1 + (u_2+y1)^T y_2 \\
   &\leq -W(x) + u_1^T y_1 + u_2^T y_2, \\
\end{split}  
\end{equation*}
which results in a classical passivity result of the feedback interconnection. Furthermore, in the case of $u_1=u_2=0$, the derivative reads as $\dot{V}(x,t) \leq -W(x)$ and all assumptions of Theorem \ref{th:invariance} are fulfilled. Then, one can obtain the domain $W(x)=0$, which is the domain where the system trajectories will be driven. \hfill $\blacksquare$
\end{pf}

\begin{rem}
Theorem \ref{th:passive} is an extension of the classical result of the interconnection between two passive systems \cite[Chapter 6]{b:Khalil2002}. Such extension covers non autonomous dynamical systems and discontinuities in both the dynamical system and the memoryless function. Theorem \ref{th:feedback} is recovered then when both systems, $H_1$ and $H_2$, fulfil the necessary smoothness conditions (locally Lipschitz around the origin) requested in the classical result. 
\end{rem}

\begin{rem}
Theorem \ref{th:passive} provides the domain in which the trajectories of the closed-loop system will converge to, in contrast to the stabilization of the origin obtained from the classical feedback interconnection of two passive systems \cite[Chapter 6]{b:Khalil2002}. 
\end{rem}

\section{Underactuated Mechanical System with Coulomb Friction}
\label{sec:Problem}

Consider an $n$-DOF underactuated mechanical system modelled as
\begin{equation}
\begin{split}
  \dot{\delta} &= \abs{v}, \\
  \dot{u} &= v, \\
  M\dot{v} &= {F}_e^{or}(u,v) - {F}_r^{or}(\delta,u,v,{p}^{or},t),
\end{split}
\label{eq:original}
\end{equation}
where $\delta \in \Re^n$, $u \in \Re^n$, $v \in \Re^n$, represent the vectors of frictional slips, displacements and velocities (slip-rates), respectively. The state $\delta(t)$ represents the accumulated slip and it can not take negative values. The term ${p}^{or} \in \Re^q$ is the vector of control inputs, where $q<n$, resulting in having more degrees of freedom (DOF) than control inputs. $M \in \Re^{n \times n}$ is the inertia matrix and the term ${F}_e^{or}(u,v) \in \Re^n$ is the vector of applied forces, which are considered to be viscoelastic forces defined as
\begin{equation}
  {F}_e^{or}(u,v)=-{K}^{or}u - {H}^{or}v,
  \label{eq:Fe}
\end{equation}
where ${K}^{or} \in \Re^{n \times n}$ is the stiffness matrix and ${H}^{or} \in \Re^{n \times n}$ is the viscosity matrix. The term ${F}_r^{or}(\delta,u,v,{p}^{or},t) =$ \\ $[{F}_{r_1}^{or}(\delta_1,u_1,v_1,{p}^{or},t),...,{F}_{r_n}^{or}(\delta_n,u_n,v_n,{p}^{or},t)]^T$ is the friction force and is written as follows
\vspace{-15pt}
{\small
\begin{equation}
\begin{split}
  &{F}_{r_i}^{or}(\delta_i,u_i,v_i,{p}^{or},t) = \\ 
  &\left \{\ \begin{split}
  {F}_i^{or}(\delta_i,v_i,{p}^{or},t) \quad &\textup{if} \quad v_i \neq 0 \\ 
  {F}_{e_i}^{or}(u_i,0) \quad &\textup{if} \quad v_i = 0 \quad \textup{and} \quad \abs{{F}_{e_i}^{or}(u_i,0)}<F_{s_i} \\ 
  F_{s_i} \sign{({F}_{e_i}^{or}(u_i,0))} \quad &\textup{if} \quad v_i = 0 \quad \textup{and} \quad \abs{{F}_{e_i}^{or}(u_i,0)} \geq F_{s_i} 
  \end{split}  \right.
\end{split}
  \label{eq:Fr}
\end{equation}}
where $i \in [1,n]$, \\ ${F}^{or}(\delta,v,{p}^{or},t)=[{F}_1^{or}(\delta_1,v_1,{p}^{or},t),...,{F}_n^{or}(\delta_n,v_n,{p}^{or},t)]^T$ is an arbitrary friction function, ${p}^{or} \in \Re^q$ is the vector of control inputs and $F_s = [F_{s_1},...,F_{s_n}]^T$ is a vector of static friction coefficients. The static friction counteracts the applied forces below a certain level and, thus, it prevents slip. 

\begin{rem}
The inertia matrix, $M$, is considered to be constant and only translational displacements are on play, \textit{i.e.}, no Coriolis/Centripetal forces are considered in this work.
\end{rem}

It is assumed that \eqref{eq:original} has an equilibrium point at $t=t^* \in \Re_{\geq 0}$. This equilibrium point is defined as $(\delta^*,u^*,v^*)$ and is described by
\begin{equation*}
\begin{split}
  \delta^* &= \delta(t^*), \quad v^* = 0, \\ 
  u^* &= u(t^*), \quad {F}_e^{or}(u^*,0) = {F}_r^{or}(\delta^*,u^*,0,p^*,t^*),
\end{split}
\end{equation*}
where $p^*=p^{or}(t^*) \in \Re^{q}$ is the vector input at the equilibrium point. It is assumed that system \eqref{eq:original} is on the verge of slip, \textit{i.e.}, $\abs{{F}_{e_i}^{or}(u_i^*,0)} \geq F_{s_i}$ for all $i \in [1,n]$ in \eqref{eq:Fr}. Therefore, we set ${F}_r^{or}(\delta^*,u^*,0,p^*,t^*) = F_s^*$, where $F_s^* \in \Re^{q}$ is the vector of friction at the equilibrium point. This is the point at which the system will be controlled. We then shift the system to this equilibrium point as follows. Let $x=[x_1,x_2,x_3]^T$ with $x_1 = \delta-\delta^*$, $x_2 = u-u^*$, $x_3 = v-v^*$ and $p={p}^{or}-p^*$. Then, we obtain
\begin{equation}
\begin{split}
  \dot{x}_1 &= \abs{x_3}, \\
  \dot{x}_2 &= x_3, \\
  \dot{x}_3 &= F_e(x_2,x_3) - M^{-1}F_r(x_1,x_2,x_3,p,t),
\end{split}
\label{eq:shift}
\end{equation}
where 
\vspace{-10pt}
{\small
\begin{equation}
\begin{split}
  F_e(x_2,x_3)&=-K x_2 - H x_3, \\
  F_r(x_1,x_2,x_3,p,t)&={F}_r^{or}(x_1+\delta^*,x_2+u^*,x_3,p+p^*,t) - F_s^*,
\end{split}
\label{eq:Fr2}
\end{equation}}

\noindent with $K=M^{-1}{K}^{or}$ and $H=M^{-1}{H}^{or}$ are defined. Recalling that the slip $\delta(t)$ is nonnegative, the new state $x_1(t)$ is nonnegative as well. The set of equilibrium points of system \eqref{eq:shift} is defined as $\Gamma(t)=\{\ x^* \in \Re^{3n} \mid x_3^*=0, {K}^{or}x_2^*=-F_r(x_1^*,x_2^*,0,p,t) \}\ $. 

The shifted system has the same form with \eqref{eq:original} except for the new term $F_s^*$ in the friction term $F_r(x_1,x_2,x_3,p,t)$. This term represents a destabilizing force due to viscoelasticity and the associated to it stored potential energy of the system, \textit{i.e.}, ${F}_e^{or}(u^*,0)=-{K}^{or}u^*=F_s^*$. From the energetic point of view, if this stored energy can not be counteracted by the friction, the system will move abruptly and a part of its stored energy will be suddenly released (instability behaviour). The prevention of such fast-slip behaviour is the main objective in this work.

\subsection{Coulomb Friction, Actuation and Passivity}

The term ${F}^{or}(\delta,v,p^{or},t)$ in \eqref{eq:Fr} is modelled as Coulomb friction \cite{b:Armstromg-Dupont-Canudas-1994,b:Olsson-Astrom-Canudas-Gafvert-Lischinsky-1998,b:Pennestri-Rossi-Salvini-Valentini-2016} and can be defined point-wise, for $i \in [1,n]$, as
\begin{equation*}
  {F}_i^{or}(\delta_i,v_i,{p}_i^{or},t) = \sign{(v_i)} \mu_i(\delta_i,\abs{v_i},t) A_i (\sigma_{n_i}-{p}_i^{or}),
\end{equation*}
where $\mu_i(\delta_i,\abs{v_i},t)$ is the friction coefficient, and $\sigma_{n_i}$ and ${p}_i^{or}$ are the normal stress and pressure applied at the surface area, $A_i$, respectively. Such friction law is a set-valued function due to the term $\sign{(v_i)}$. The term ${p}_i^{or}$ could be seen as an input to modify the friction: when the pressure ${p}_i^{or}$ increases, the friction ${F}_i^{or}(\delta_i,v_i,{p}_i^{or},t)$ decreases, and vice versa. Nevertheless, in real applications is not feasible to change the pressure at every point, \textit{i.e.}, it is not possible to change the value of every ${p}_i^{or}$ independently.

A way to relate the point-wise pressure ${p}_i^{or}$ with the pressure input ${p}^{or}$ in system \eqref{eq:original} is through a relation matrix\footnote{A relation (logical, boolean or binary) matrix is a matrix with only entries of zeros or ones \cite{b:bams/1183523748}.} $C_p \in \Re^{n \times q}$, \textit{i.e.}, $[{p}_1^{or},...,{p}_n^{or}]^T=C_p {p}^{or}$. This allows to reduce the number of inputs of the system by paying the price of underactuation. Notice that the matrix $C_p$ has to be full rank and to have nonzero rows. These conditions are justified by the physics of the problem and they are related to the controllability of the system.

Therefore, the Coulomb friction is defined for the whole system as
\begin{equation}
  {F}^{or}(\delta,v,{p}^{or},t) = \sign{(v)} \mu(\delta,\abs{v},t) A (\sigma_n-C_p {p}^{or}),
  \label{eq:Coulomb}
\end{equation}
where the term $\mu(\delta,\abs{v},t) \in \Re^{n \times n}$ is defined as $\mu(\delta,\abs{v},t)= \textup{diag}\left[\mu_1(\delta_1,\abs{v_1},t), ..., \mu_n(\delta_n,\abs{v_n},t) \right]$, where $\mu_i(\delta_i,\abs{v_i},t)$, with $i \in [1,n]$, are friction coefficients. $A$ is the surface area of the frictional interface, defined as $A=\textup{diag}\left[A_1, ..., A_n \right]$. The effective stress $\sigma_n-C_p {p}^{or}$ is defined, with $\sigma_n \in \Re^n$ as a vector of normal stresses ($\sigma_n=[\sigma_{n_1}, ..., \sigma_{n_n}]^T$), and the matrix $C_p \in \Re^{n \times q}$ as the relation matrix, ruling how the control input, ${p}^{or}$, influences the system.

\begin{figure*}[ht!]
  \centering 
  \includegraphics[width=5.7cm,height=3cm]{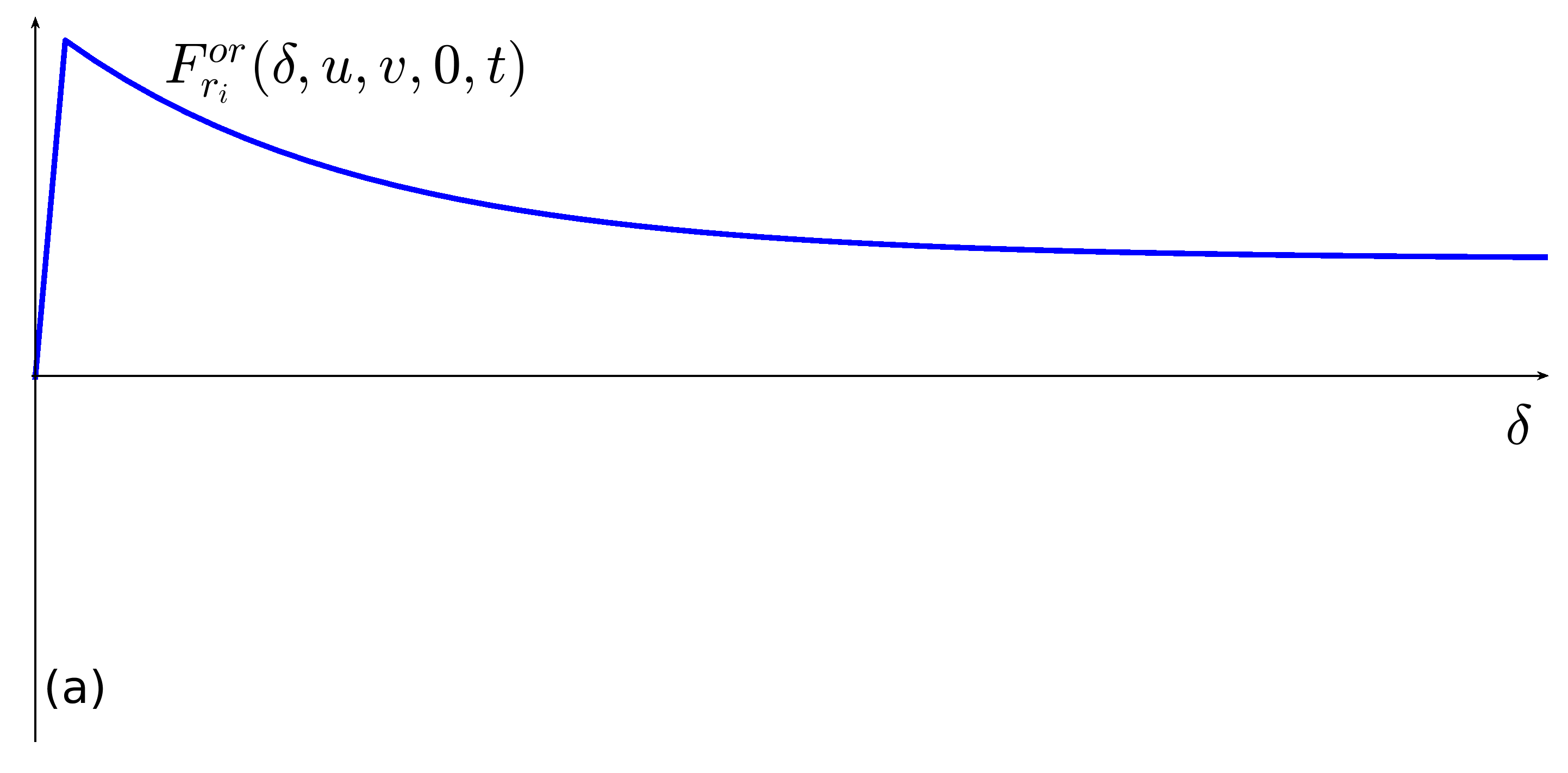}
  \includegraphics[width=5.7cm,height=3cm]{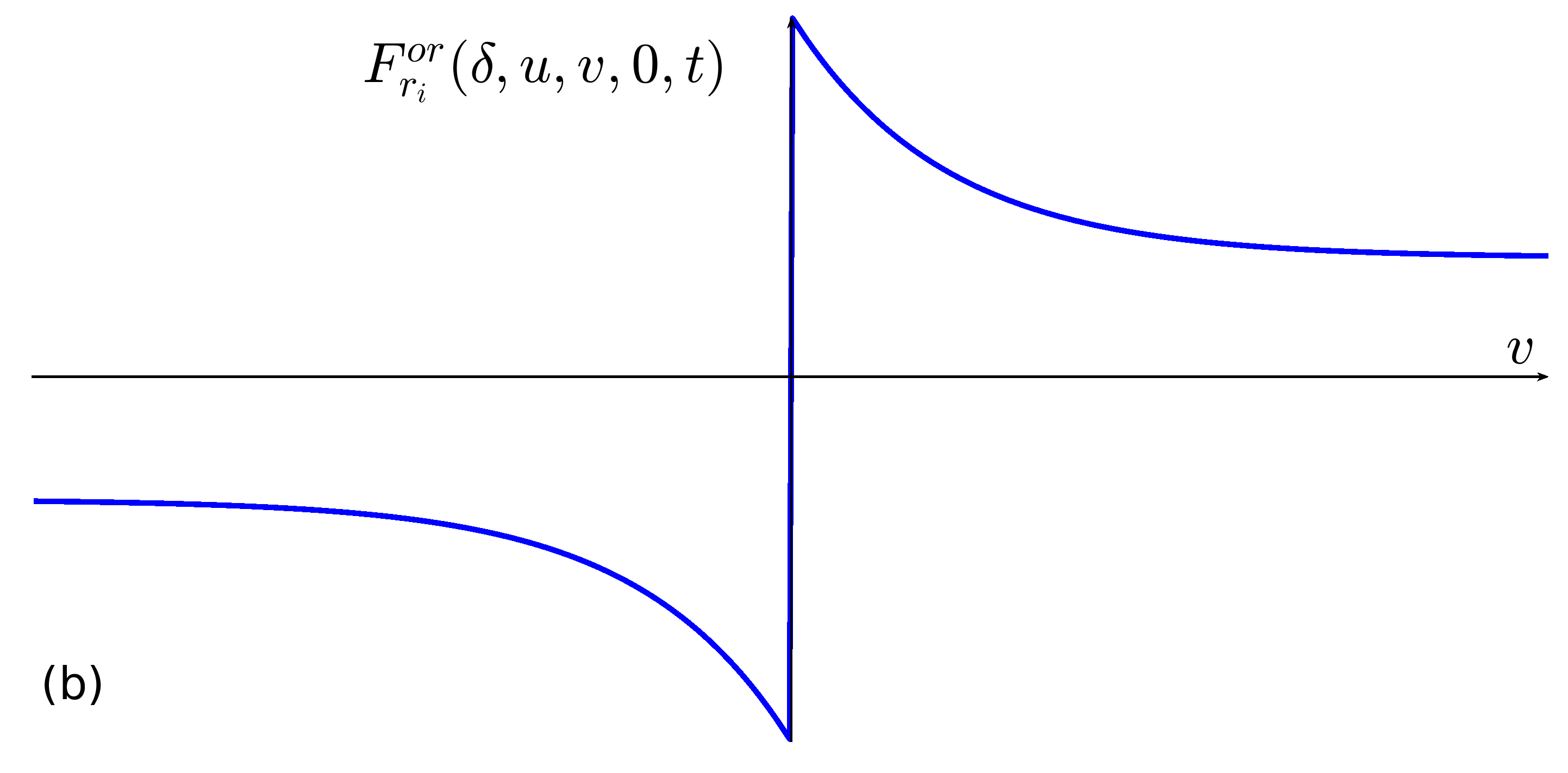}\\
  \includegraphics[width=5.7cm,height=3cm]{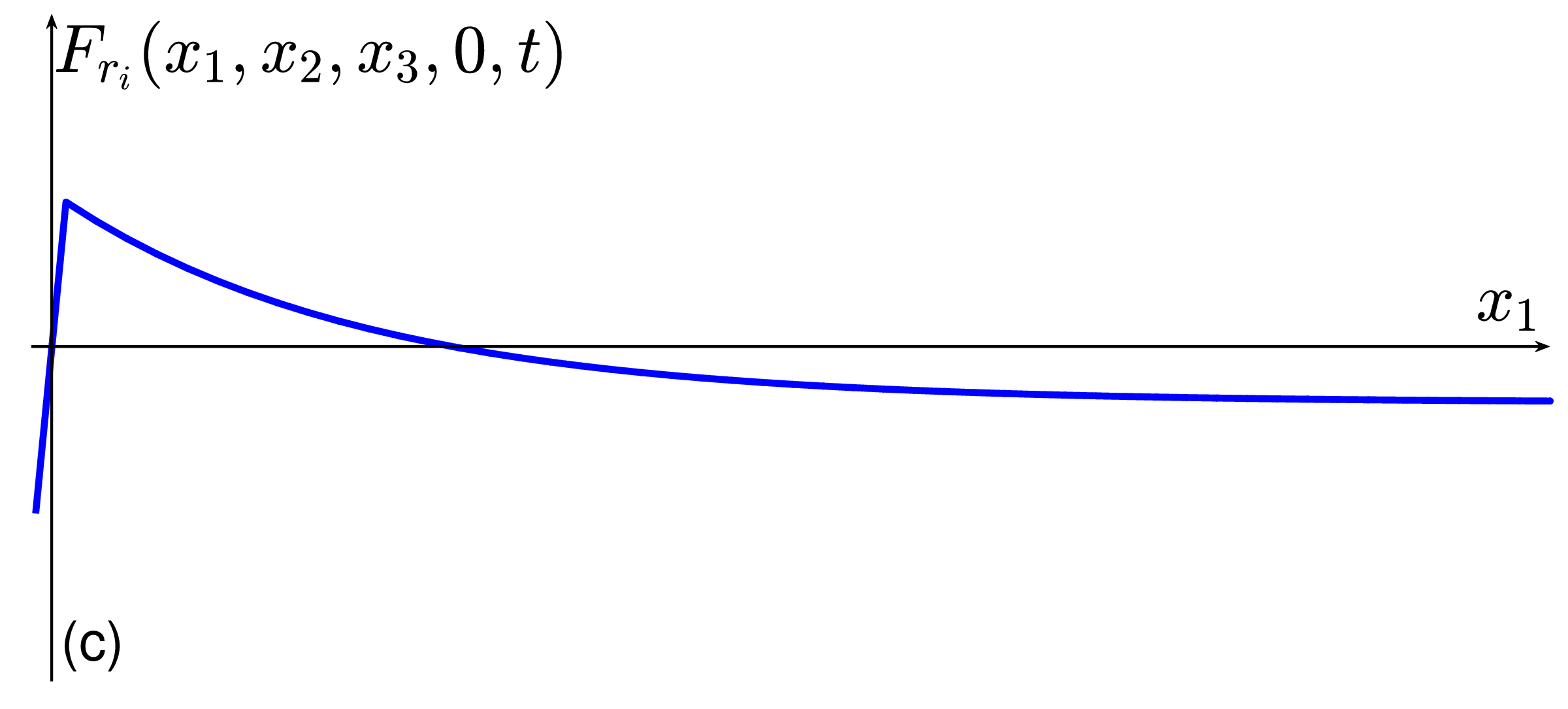}
  \includegraphics[width=5.7cm,height=3cm]{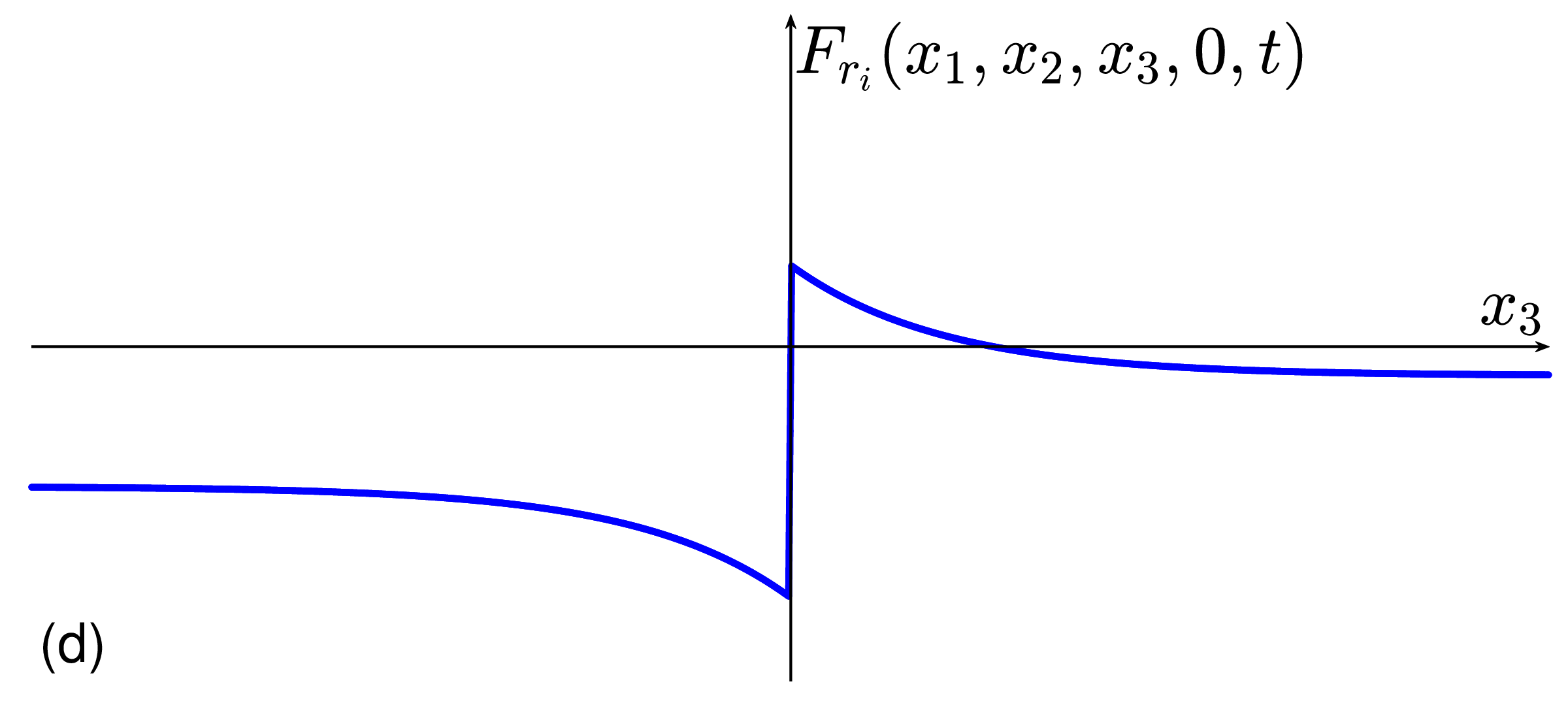}
  \caption{(a)-(b): Schematic representation of a component of ${F}_r^{or}(\delta,u,v,0,t)$, showing how it is passive with respect to the input $[\delta^T,v^T]^T$.  (c)-(d): Loss of passivity due to the addition of the loading term $F_s^*$ resulting in the new term $F_r(x_1,x_2,x_3,0,t)$.}
  \label{fig:Fr}
\end{figure*}

A schematic plot of ${F}_r^{or}(\delta,u,v,{p}^{or},t)$ defined as \eqref{eq:Fr}, \eqref{eq:Coulomb}, with ${p}^{or}=0$ is shown in Fig. \ref{fig:Fr}. The function ${h}^{or}(\delta,u,v,0,t)=[0_{1 \times n}, {F}_r^{or}(\delta,u,v,0,t)^T]^T$, ${h}^{or}:\Re^n \times \Re^n \times \Re^n \times \Re^q \times \Re_{\geq 0} \rightarrow \Re^{2n}$ is passive belonging to the sector $[0,\infty]$, with $[\delta^T,v^T]^T$ as input\footnote{See \cite[Chapter 6]{b:Khalil2002} for more details about sector definition in passivity.}.

According to \eqref{eq:Fr2}, $F_s^*$ translates ${F}_r^{or}(\delta,u,v,0,t)$ in the new system \eqref{eq:shift} (see Fig. \ref{fig:Fr}(c-d)). As a result, the passivity property of the output $h(x_1,x_2,x_3,0,t)=[0_{n}, F_r(x_1,x_2,x_3,0,t)]$, $h:\Re^n \times \Re^n \times \Re^n \times \Re^q \times \Re_{\geq 0} \rightarrow \Re^{2n}$ is lost. 

The new shifted friction term \eqref{eq:Coulomb} can be written as
\begin{equation}
\begin{split}
  F(x_1,x_3,p,t) &= g(x_1,x_3,t)-b(x_1,x_3,t)C_p p, \\
  g(x_1,x_3,t) &= \sign{(x_3)} \mu(x_1+\delta_0,\abs{x_3},t) A \sigma_n^{\prime}-F_s^*,\\
  b(x_1,x_3,t) &= \sign{(x_3)} \mu(x_1+\delta_0,\abs{x_3},t) A,
\end{split}
  \label{eq:Coulomb2}
\end{equation}
where $\sigma_n^{\prime}=\sigma_n-C_p p_0$ is a vector of constant values. If the control input $p \in \Re^q$ is taken into account in \eqref{eq:Coulomb2}, the original passivity property could be recovered in the shifted friction term and a stability result for system \eqref{eq:shift} could be obtained.

\subsection{Control Objectives}

The control objectives are stated as follows:
\begin{enumerate}
  \item To design the control $p$ in \eqref{eq:shift}, \eqref{eq:Fr2} and \eqref{eq:Coulomb2} such that the output $y_2=[-\abs{x_3}^T,F_r(x_1,x_2,x_3,p,t)^T]^T$ to become passive. 
  \item To design an integral action to the latter control law, obtaining a reference tracking over the output error
  \begin{equation}
    y_t = C_t(r_3-x_3),
    \label{eq:outt}
  \end{equation}   
  where $C_t \in \Re^{q \times n}$ is a matrix to be defined and $r_3\in \Re^n$ is a vector of constant velocity references.
  \item Considering dynamics in the input $p$ (actuator dynamics) as
  \begin{equation}
    \dot{p} = C_h(p_\infty-p), 
    \label{eq:actdyn}
  \end{equation}
  where $C_h \in \Re^{q \times q}$, to design the new control input $p_\infty \in \Re^q$ capable to reproduce the same results as the ones obtained in objectives 2 and 3.
\end{enumerate}
The above mentioned control objectives correspond to three distinct design steps, whose role is explained as follows. The first step allows the friction to recover the lost passivity, whereas the second step allows to release the stored energy of the system slowly, by choosing a small velocity reference $r_3$. The final step accounts for the dynamics of the actuator and allows the design of the real control input $p_\infty$.

\begin{figure}[ht!]
  \centering 
  \includegraphics[width=8.5cm,height=5.5cm]{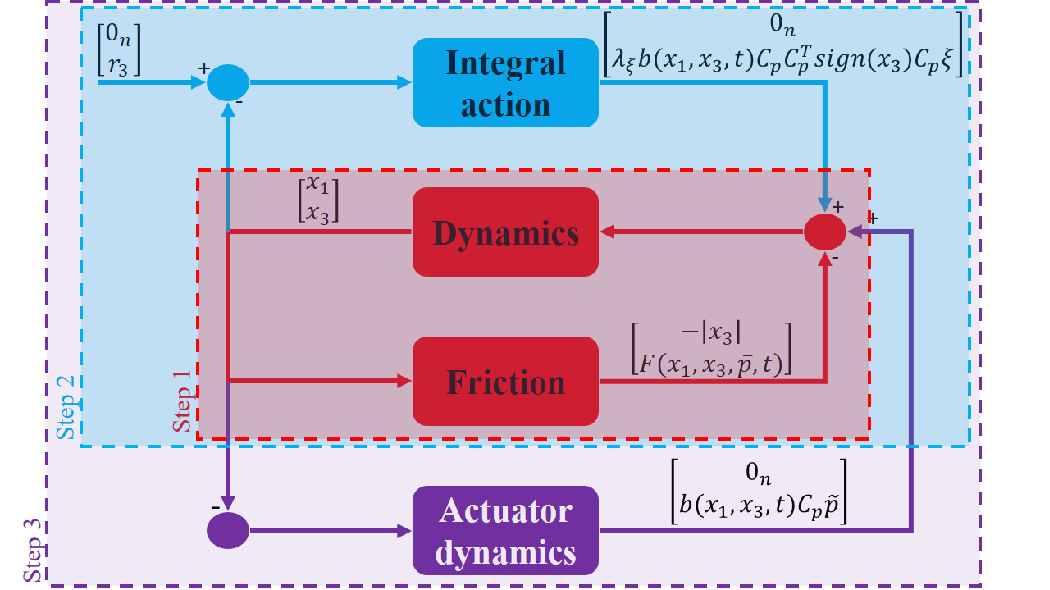}
  \caption{Steps of the passivity-based design.}
  \label{fig:Fulldiagram}
\end{figure}

\begin{figure}[ht!]
  \centering 
  \includegraphics[width=8.5cm,height=2.5cm]{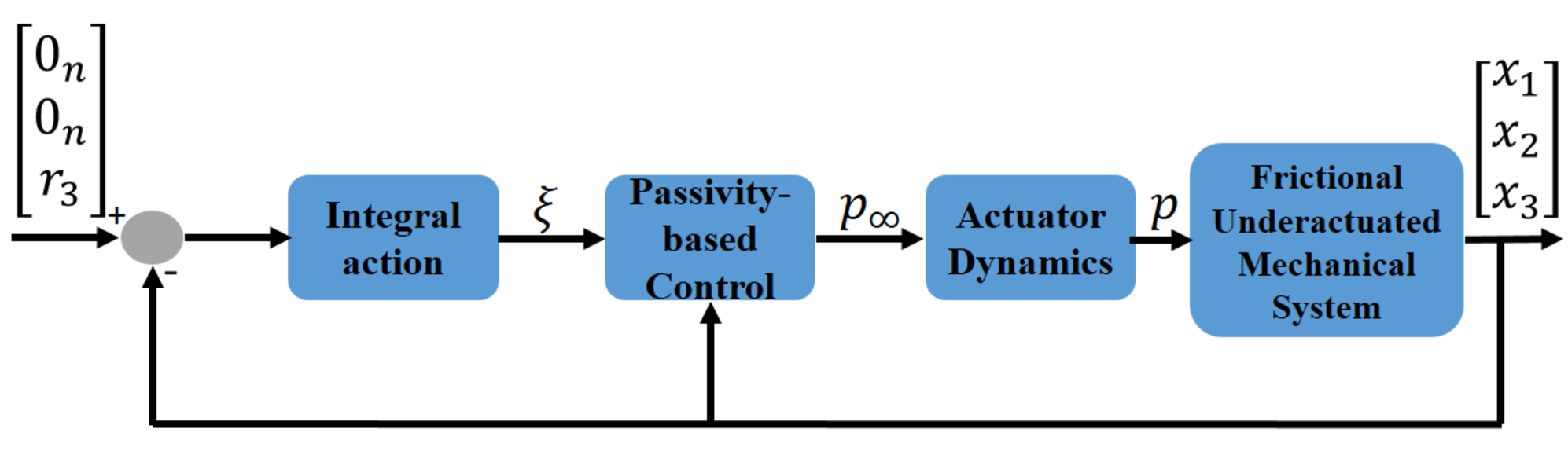}
  \caption{Closed loop system.}
  \label{fig:Fulldiagram2}
\end{figure}

A block diagram of the full passivity-based control design is shown in Fig. \ref{fig:Fulldiagram} and the closed-loop system is illustrated in Fig. \ref{fig:Fulldiagram2}. The description of every part of this design is explained in the following sections. 

The control design will be performed under the next following minimal assumptions for system \eqref{eq:shift}:
\begin{assum}\label{A1}
The initial condition of system \eqref{eq:shift} will be the origin: $x_1(0)=x_2(0)=x_3(0)=0$.
\end{assum}
\begin{assum}\label{A2}
The inertia matrix $M$ is symmetric and positive definite, \textit{i.e.}, $M=M^T>0_{n\times n}$.
\end{assum}
\begin{assum}\label{A3}
Matrices ${K}^{or},K,{H}^{or},H,C_h$ are positive definite. Furthermore, ${K}^{or}$, ${H}^{or}$ and $C_h$ are symmetric matrices.
\end{assum}
\begin{assum}\label{A4}
The friction coefficient satisfies $\min \{ \mu(x_{1}+\delta_{0},\abs{x_{3}},t)A \} = \mu_{min} > 0$. Furthermore, $\mu_{min}$ is a known constant.
\end{assum}
\begin{assum}\label{A5}
The function $h=[0_{1\times n},F_r(x_1,x_2,x_3,0,t)^T]^T$ belongs to the sector $[L_{F_r},\infty]$, with \\ $L_{F_r}=\left[\begin{array}{cc}
 0_{n \times n} & 0_{n \times n} \\ 
 -l_\delta \sign{(x_3)} & -l_v I_{n}
 \end{array}  \right]$, input $[x_1^T,x_3^T]^T$ and $l_\delta,l_v>0$ assumed to be known constants (see Fig. \ref{fig:Fr}).
\end{assum}
\begin{assum}\label{A6}
Relation matrix $C_p$ in \eqref{eq:Coulomb2} have full rank, has nonzero rows and it is known.
\end{assum}

\begin{rem}
Assumptions \ref{A2}-\ref{A4} are fulfilled commonly in mechanical systems. Furthermore, $\mu_{min}$ always exist due to thermodynamics (energy conservation).
\end{rem}

\begin{rem}
Assumption \ref{A5} is physically justified by empirical frictional laws that are always bounded (see \cite{b:Armstromg-Dupont-Canudas-1994,b:Olsson-Astrom-Canudas-Gafvert-Lischinsky-1998,b:https://doi.org/10.1029/2021JB023410})
\end{rem}
  
\section{Passivity-based Control Design of Underactuated Frictional Systems}
\label{sec:Control}

\subsection{Stabilization of the Frictional System}

When the system is in motion, \textit{i.e.} $x_3 \neq 0$, the frictional term $F_r(x_1,x_2,x_3,p,t)$ in \eqref{eq:Fr}, \eqref{eq:Fr2}, turns into $F(x_1,x_3,p,t)$ described by \eqref{eq:Coulomb2}, which will be considered in the subsequent analysis. 

\begin{figure}[ht!]
  \centering 
  \includegraphics[width=6cm,height=3.5cm]{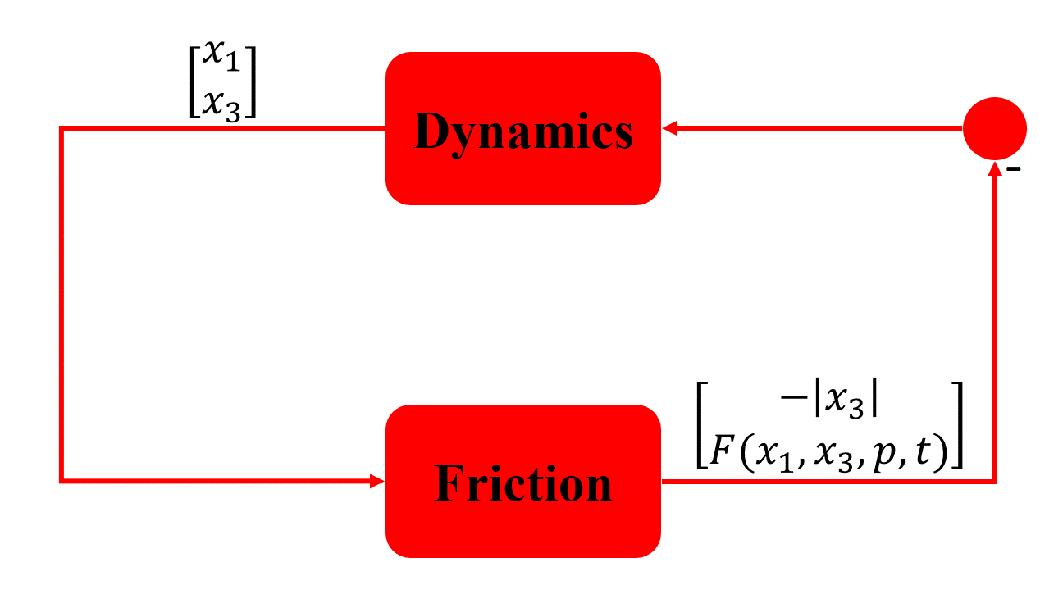}
  \caption{Control design: Step 1.}
  \label{fig:Diagram1}
\end{figure}

The feedback interconnection between a mechanical system and a frictional term, \textit{i.e.} system \eqref{eq:shift}, will be analysed using Theorem \ref{th:passive}. Such interconnection can be seen in Fig. \ref{fig:Diagram1} and is the same as in Fig. \ref{fig:Feedback}, where system $H_1$ is defined as 
$\dot{x}_1 = e_{11}$, $\dot{x}_2 = x_3$, $\dot{x}_3 = F_e(x_2,x_3)+M^{-1}e_{21}$ ($e_{11}=\abs{x_3}$ and $e_{21}=-F(x_1,x_3,p,t)$), with $u_1=0_{2 n}$, $e_1=[e_{11}^T,e_{21}^T]^T$, and $y_1=[x_1^T,x_3^T]^T$. The system $H_2$ is defined as the memoryless function $y_2=[-\abs{x_3},F(x_1,x_3,p,t)]$ with $u_2=0_{2 n}$, and $e_2=[x_1^T,x_3^T]^T$.

The next Lemma will show the passivity property of the frictional term, $F(x_1,x_3,p,t)$, when the control input, $p$, is now taking into consideration.
\begin{lem}
The passivity map $e_2^T y_2$ is passive, \textit{i.e.}, $e_2^T y_2 \geq 0$, if the control input $p$ is defined as
\begin{equation}
  p(x_1,x_3) = - \lambda_\delta C_p^T x_1 - \lambda_v C_p^T \abs{x_3},
  \label{eq:p}
\end{equation}
with control gains, $\lambda_\delta,\lambda_v$, satisfying
\begin{equation}
  \lambda_\delta > \frac{l_\delta+1}{\mu_{min}}, \quad \lambda_v > \frac{l_v}{\mu_{min}}.
  \label{eq:cond1}
\end{equation}
\end{lem}

\begin{pf}
According to the definition of a sector in \cite[Chapter 6]{b:Khalil2002} and eqs. \eqref{eq:Fr2}, \eqref{eq:Coulomb2}, Assumption \ref{A5} leads to
 \begin{equation}
   x_3^T g(x_1,x_3,t) \geq -l_\delta \abs{x_3}^T x_1 - l_v x_3^T x_3.
   \label{eq:gsector}
 \end{equation}
Therefore, the passivity map, $e_2^T y_2$, reads
\begin{equation*}
\begin{split}
  e_2^T y_2 &= - x_1^T \abs{x_3} +  x_3^T g(x_1,x_3,t) - x_3^T b(x_1,x_3,t)C_p p \\
  &\geq -(l_\delta+1) \abs{x_3}^T x_1 - l_v x_3^T x_3 \\ &\quad- x_3^T \sign{(x_3)}^T \mu(x_1+\delta_0,\abs{x_3},t) A C_p p \\
  &\geq -(l_\delta+1) \abs{x_3}^T x_1 - l_v x_3^T x_3 - \abs{x_3}^T \mu_{min} C_p p,
\end{split}
\end{equation*}
where the sector condition \eqref{eq:gsector} and Assumption \ref{A4} for the term $\mu(x_1+\delta_0,\abs{x_3},t)A$ have been used.

Selecting the control input $p$ as \eqref{eq:p}, where $\lambda_\delta,\lambda_v$ are constants to be designed, the passivity map becomes
\begin{equation*}
\begin{split}
  e_2^T y_2 &\geq \abs{x_3}^T \left( \frac{\mu_{min}\lambda_\delta}{l_\delta+1} C_p C_p^T- I_{n \times n} \right) x_1 \\ &\quad + \abs{x_3}^T \left( \frac{\mu_{min}\lambda_v}{l_v} C_p C_p^T- I_{n \times n} \right) \abs{x_3}.
\end{split}
\end{equation*}

Due to Assumption \ref{A6}, the product $C_p C_p^T$ is nonnegative (\textit{i.e.}, all its elements are nonnegative) and the elements of the diagonal are greater or equal to one. Therefore, last expression results to be passive, \textit{i.e.} $e_2^T y_2 \geq 0$, if the controller gains are chosen as \eqref{eq:cond1}.
\hfill $\blacksquare$.
\end{pf}

Notice that the designed control input $p$ in \eqref{eq:p} injects passivity into the shifted friction term $g(x_1,x_3,t)$. However, a strict passivity condition can not be obtained due to the underactuation nature of the system, \textit{i.e.} $C_p C_p^T \geq 0$. Nevertheless, this is not a critical condition for the stability result stated in the next Theorem.

\begin{thm}
Every bounded solution $x(t)$ of system \eqref{eq:shift} approaches to the domain $E= \{\ x \in \Re^{3n} \mid x_3=0 \}\ $ as $t \rightarrow \infty$, if the control input $p(x_1,x_3)$ is defined as in \eqref{eq:p} and \eqref{eq:cond1}.
\label{th1}
\end{thm}

\begin{pf}
Consider the positive definite function
\begin{equation*}
  V(x) = \frac{1}{2}x_1^T x_1 +  \frac{1}{2} x_2^T {K}^{or} x_2 + \frac{1}{2} x_3^T M x_3,
\end{equation*}
and its time derivative along the trajectories of system \eqref{eq:shift} as
\begin{equation*}
\begin{split}
  \dot{V} &= x_1^T \abs{x_3} + \frac{1}{2}x_3^T {K}^{or} x_2 +\frac{1}{2} x_2^T {K}^{or} x_3 \\ 
  &\quad + \frac{1}{2} \left[ -Kx_2 - Hx_3 - M^{-1}F(x_1,x_3,p,t) \right] ^T M x_3 \\
  &\quad + \frac{1}{2}x_3^T M \left[ -Kx_2 - Hx_3 - M^{-1}F(x_1,x_3,p,t) \right] \\
  &= x_1^T \abs{x_3} - x_3^T F(x_1,x_3,p,t)- x_3^T {H}^{or} x_3 \\
  &=e_1^T y_1 - x_3^T {H}^{or} x_3, 
\end{split}
\end{equation*}
which results to be passive due to Assumption \ref{A3}. If the controller gains are chosen as in \eqref{eq:cond1}, the frictional term is passive and $e_2^T y_2 \geq 0$. Consequently, using Theorem \ref{th:passive}, every bounded solution $x(t)$ of system \eqref{eq:shift} (the feedback interconnection between two passive systems with $u_1=u_2=0_{2 n}$) converges to the domain $E= \{\ x \in \Re^{3n} \mid x_3=0 \}\ $ as $t \rightarrow \infty$. \hfill $\blacksquare$
\end{pf}

Notice that the above mentioned domain is bounded, given the first two equations of system \eqref{eq:shift}, \textit{i.e.}, $x_1(t)$ and $x_2(t)$ will become constant and, therefore, they are bounded.

The presented stability result is not as strong as the asymptotic (or exponential) stability of the system origin. Nevertheless, recalling the definition of system \eqref{eq:shift}, it results in an increasing evolution of the state $x_1(t)$ and the impossibility of returning it to the origin once it has started to evolve. Therefore, the obtained stability result is the best that one can obtain for these kind of frictional systems.

\subsection{Regulation via Integral Control}

The stability result of the previous section prohibits the abrupt release of the stored energy of the system by immobilizing it (convergence of system trajectories to zero velocities). However, the energy is still trapped into the system, requiring its stabilization continuously. For this purpose, tracking will be performed in order to allow the system to follow a constant (small) reference, $r_3$, and release the stored energy with small velocities. In other words, the small reference, $r_3$, will bring the system to another state of lower energy. This will be achieved by interconnecting the underactuated mechanical system with an integral extension of the tracking error.

Considering the new integral term
\begin{equation}
  \dot{\xi} = y_t = C_t(r_3-x_3),
  \label{eq:integral}
\end{equation}
where $\xi \in \Re^q$ and $y_t$ is the error variable defined in \eqref{eq:outt}. Following a classical integral design (see \textit{e.g.} \cite[Chapter 12]{b:Khalil2002}), let us define the regulation error variables as
\begin{equation}
\begin{split}
  x_{i_e}(t) &= x_i(t)-x_i(\infty), \quad p_e(t) = p(t)-p(\infty), \\ \xi_e(t) &= \xi(t) - \xi(\infty), 
\end{split}
  \label{eq:trerror}
\end{equation}
where $i = 1,2,3$ and $x_i(\infty),p(\infty),\xi(\infty)$ are the steady state values of the states, the control input and the integral action, respectively.

The error dynamics is written as
\begin{align}
  \dot{x}_{1_e} &= \abs{x_3}-\abs{x_3(\infty)} \nonumber \\ 
  &= \abs{x_{3_e}+x_3(\infty)}-\abs{x_3(\infty)} \leq \abs{x_{3_e}}, \label{eq:trdyna}\\
  \dot{x}_{2_e} &= x_{3_e},  \\
  \dot{x}_{3_e} &= F_e(x_{2_e},x_{3_e}) - M^{-1}\Delta F(x_{1_e},x_{3_e},p_e,t), \label{eq:trdync}\\
  \dot{\xi}_e &= -C_t x_{3_e} \label{eq:trdynd},
\end{align}
due to the fact that $r_3=r_3(\infty)$, because $r_3$ is a vector of constant references. The new nonlinear function $\Delta F(x_{1_e},x_{3_e},p_e,t)$ is defined as
\begin{equation}
\begin{split}
  &\Delta F(x_{1_e},x_{3_e},p_e,t) \\&= F(x_{1},x_{3},p,t)-F(x_{1}(\infty),x_{3}(\infty),p(\infty),t)\\
  &= F(x_{1_e}+x_{1}(\infty),x_{3_e}+x_{3}(\infty),p_e+p(\infty),t)\\ &\quad-F(x_{1}(\infty),x_{3}(\infty),p(\infty),t),
\end{split}
\end{equation}
which has the same characteristics as the term \eqref{eq:Coulomb2}. Consequently, the control input $p_e$ of the error dynamics in \eqref{eq:trdyna}-\eqref{eq:trdynd} can be designed as
\begin{equation}
\begin{split}
  p_e(x_{1_e},x_{3_e},\xi_e) &= - \lambda_\delta C_p^T x_{1_e} - \lambda_v C_p^T \abs{x_{3_e}} \\ &\quad + \lambda_\xi C_p^T \sign{(x_{3_e}+x_3(\infty))} C_p \xi_e,
\end{split}
  \label{eq:pte}
\end{equation}
where $\lambda_\xi \in \Re_{>0}$ is a gain to be designed. The first two terms of the latter control are designed to stabilize the mechanical system, equivalent to the system as in Theorem \ref{th1}, while the new term includes the integral action to perform the regulation.

\begin{figure}[ht!]
  \centering 
  \includegraphics[width=7.5cm,height=4.3cm]{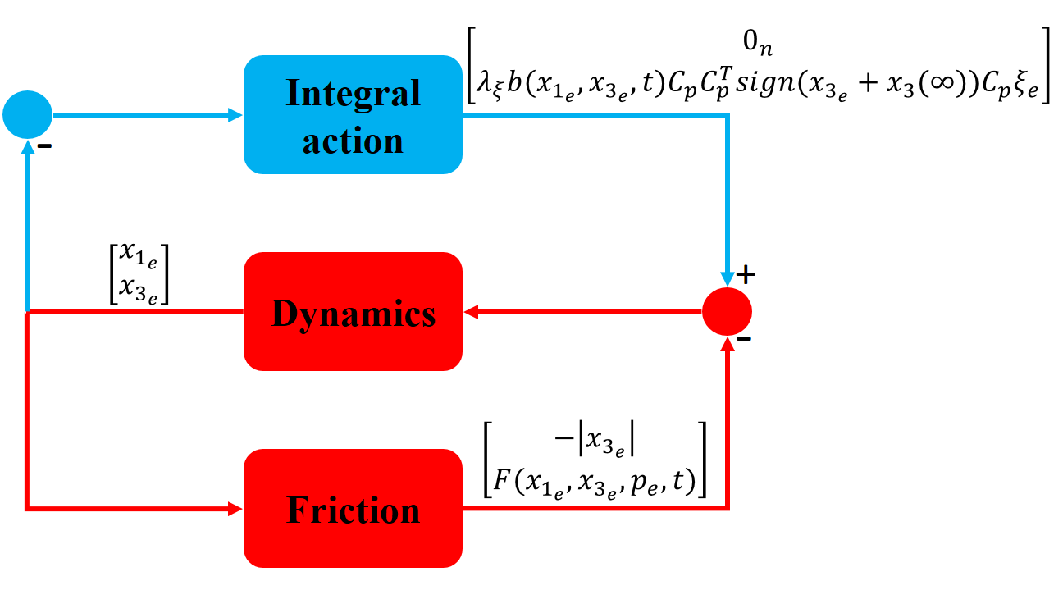}
  \caption{Control design: Step 2.}
  \label{fig:Diagram2}
\end{figure}

The interconnection of the mechanical system and the integral action is shown in Fig. \ref{fig:Diagram2}. Control \eqref{eq:pte} interconnects the two systems as in Fig. \ref{fig:Feedback}: system $H_1$ is defined as \eqref{eq:trdynd} with $u_1=0_{2 n}$, $e_1=[-x_{1_e}^T,-x_{3_e}^T]^T$, and \\ \vspace{-20pt} {\small$y_1=[0_{1 \times n},(\lambda_\xi b(x_{1_e},x_{3_e},t) C_p C_p^T \sign{(x_{3_e}+x_3(\infty))} C_p \xi_e)^T]^T$}, and system $H_2$ is defined as \eqref{eq:trdyna}-\eqref{eq:trdync} with $u_2=0_{2 n}$, \\ \vspace{-20pt} {\small$e_2=[0_{1 \times n},(\lambda_\xi b(x_{1_e},x_{3_e},t) C_p C_p^T \sign{(x_{3_e}+x_3(\infty))} C_p \xi_e)^T]^T$}, and $y_2=[x_{1_e}^T,x_{3_e}^T]^T$.

The next Theorem for the regulation solution holds. 
\begin{thm}
Every bounded solution $(x(t),\xi(t))$ of the closed-loop system \eqref{eq:shift}, \eqref{eq:integral} approaches to the domain $E_t= \{\ (x,\xi) \in \Re^{3n}\times \Re^q \mid C_t(r_3-x_3) =0_{q} \}\ $ as $t \rightarrow \infty$ if the control input $p(x_1,x_3,\xi)$ is defined as 
\begin{equation}
\begin{split}
  p(x_1,x_{3},\xi) &= - \lambda_\delta C_p^T x_{1} - \lambda_v C_p^T \abs{x_{3}-r_3} - \lambda_v C_p^T \abs{r_3} \\ &\quad + \lambda_\xi C_p^T \sign{(x_{3})} C_p \xi,
\end{split}
  \label{eq:pt}
\end{equation}
satisfying the condition \eqref{eq:cond1}, $\lambda_\xi >0$, and 
\begin{equation}
  C_t = (C_p^T C_p)^{-1}C_p^T.
  \label{eq:Ct}
\end{equation}
\label{th2}
\end{thm}
\vspace{-30pt}
\begin{pf}
From the previous stability result, we know that the system \eqref{eq:shift} is passive, \textit{i.e.}, $e_2^T y_2 \geq \dot{V} + x_3^T {H}^{or} x_3$, if $p$ is designed as \eqref{eq:p}, \eqref{eq:cond1}. Such result can be inherited to the equivalent system \eqref{eq:trdyna}-\eqref{eq:trdync}. Thus, now the passivity property must be studied in system \eqref{eq:trdynd}.

Let us study first the passivity map of the output $L(C_p \xi_e)=\lambda_\xi b(x_{1_e},x_{3_e},t) C_p C_p^T \sign{(x_{3_e}+x_3(\infty))} C_p \xi_e$ with input $C_p \xi_e$, resulting in
\vspace{-20pt}
{\small
\begin{equation*}
\begin{split}
  \xi_e^T C_p^T L &= \xi_e^T C_p^T \lambda_\xi b(x_{1_e},x_{3_e},t) C_p C_p^T \sign{(x_{3_e}+x_3(\infty))} C_p \xi_e\\
  &= \xi_e^T C_p^T \lambda_\xi A \mu(x_1+\delta_0,\abs{x_3},t) \sign{(x_{3_e}+x_3(\infty))} C_p C_p^T \\ &\quad \times \sign{(x_{3_e}+x_3(\infty))} C_p \xi_e\\
  &\geq \xi_e^T C_p^T \lambda_\xi \mu_{min} \sign{(x_{3_e}+x_3(\infty))} C_p C_p^T \\ &\quad \times \sign{(x_{3_e}+x_3(\infty))} C_p \xi_e \geq 0,
\end{split}
\end{equation*}}\\
\noindent where the definition of $b(x_1,x_3,t)$ in \eqref{eq:Coulomb2} and the Assumption \ref{A4} were used. Clearly, this output is passive.

Defining the storage function $V_\xi=\int_{0}^{C_p \xi_e} L(\sigma) d\sigma$ for the system \eqref{eq:integral}. Such function is positive semidefinite due to the passive property of the output $L(C_p \xi_e)$ and its derivative reads as
\begin{equation*}
\begin{split}
  \dot{V}_\xi &= L(C_p \xi_e)^T C_p \dot{\xi_e}= -x_{3_e}^T (C_t C_p)^T L(C_p \xi_e)= e_1^T y_1,
\end{split}
\end{equation*}
if $C_t$ is the left pseudoinverse matrix of $C_p$, \textit{i.e.}, $C_t$ is defined as in \eqref{eq:Ct}. Consequently, matrix $C_t$ is full rank due to Assumption \ref{A6}. The last expression shows how the integral system is passive. Therefore, the feedback connection between the two systems will be passive.

Consequently, using Theorem \ref{th:passive}, every bounded solution $(x_e(t),\xi_e(t))$ of system \eqref{eq:trdyna}-\eqref{eq:trdynd} converges to the domain $E_t= \{\ (x_e,\xi_e) \in \Re^{3n}\times \Re^q \mid x_{3e}^T {H}^{or} x_{3e} =0_n \}\ $ as $t \rightarrow \infty$. In order to obtain the domain in the original states, the error $x_{3_e}$ must fulfil the equation $C_tx_{3_e}=C_t(r_3-x_3)$ obtaining the domain $E_t$ described in Theorem \ref{th2}. 

Finally, the original control $p$ results from the control \eqref{eq:pte} and the definition of errors \eqref{eq:trerror}
\begin{equation*}
\begin{split}
  p-p(\infty) &= - \lambda_\delta C_p^T x_{1} - \lambda_v C_p^T \abs{x_{3}-r_3} \\ 
  &\quad + \lambda_\xi C_p^T \sign{(x_{3})} C_p \xi + \lambda_\delta C_p^T x_{1}(\infty) \\
  &\quad - \lambda_\xi C_p^T \sign{(x_{3})} C_p \xi(\infty),
\end{split}
\end{equation*}
where one can obtain expression \eqref{eq:pt} by replacing the steady state control $p(\infty)=-\lambda_\delta C_p^T x_{1}(\infty) - \lambda_v C_p^T \abs{x_{3}(\infty)} + \lambda_\xi C_p^T \sign{(x_{3})} C_p \xi(\infty)$. \hfill $\blacksquare$
\end{pf} 

\subsection{Actuator Dynamics}

So far, the designed control \eqref{eq:pt} is able to either drive the system \eqref{eq:shift} states to a given domain $E= \{\ x \in \Re^{3n} \mid x_3=0 \}\ $ as $t \rightarrow \infty$, if $r_3=\lambda_\xi=0$, or to perform a tracking over a given velocity constant reference if $r_3 \neq 0$ and $\lambda_\xi > 0$. If now an actuator dynamics like \eqref{eq:actdyn} is considered in the model, $p_\infty$ is the new control input to be designed. For this purpose, consider the control \eqref{eq:pt} as nominal control $\bar{p}$, \textit{i.e.},
\begin{equation}
\begin{split}
  \bar{p}(x_1,x_{3},\xi) &= - \lambda_\delta C_p^T x_{1} - \lambda_v C_p^T \abs{x_{3}-r_3} - \lambda_v C_p^T \abs{r_3} \\
  &\quad + \lambda_\xi C_p^T \sign{(x_{3})} C_p \xi.
\end{split}
  \label{eq:ptnom}
\end{equation}
Then, one can get the nominal control $\bar{p}_\infty$ from \eqref{eq:shift}, \eqref{eq:outt}, \eqref{eq:actdyn}, \eqref{eq:integral} and \eqref{eq:pt} as
\begin{equation}
\begin{split}
  \bar{p}_\infty &= C_h^{-1} \dot{\bar{p}} + \bar{p}, \\
  &= -\lambda_\delta C_p^T x_1 - \lambda_\delta C_h^{-1}C_p^T \abs{x_3} - \lambda_v C_p^T \abs{x_3-r_3} \\ 
  &\quad - \lambda_v C_h^{-1} C_p^T \sign{(x_3-r_3)} (\dot{x}_3-\dot{r}_3) - \lambda_v C_p^T \abs{r_3} \\ 
  &\quad - \lambda_v C_h^{-1}C_p^T \sign{(r_3)}\dot{r}_3 + \lambda_\xi C_p^T \sign{(x_3)} C_p \xi \\
  &\quad + \lambda_\xi C_h^{-1} C_p^T \sign{(x_3)} C_p C_t (r_3-x_3).
\end{split}
\label{eq:pinfnom}
\end{equation}
The time derivative of $\sign{(x_3)}$ is equal to zero because we are studying the case when the system is in motion ($x_3 \neq 0$).

In order to obtain the control $p_\infty$ able to reproduce the nominal control \eqref{eq:pinfnom}, let us define the next error variables
\begin{equation}
\begin{split}
  \tilde{p} = p - \bar{p}, \quad
  \tilde{p}_\infty = p_\infty - \bar{p}_\infty, 
\end{split}
\end{equation} 
leading to the error dynamics from \eqref{eq:shift}, \eqref{eq:actdyn}, \eqref{eq:integral} and \eqref{eq:pinfnom} as
\begin{align}
  \dot{x}_1 &= \abs{x_3}, \label{eq:errora}\\
  \dot{x}_2 &= x_3,  \\
  \dot{x}_3 &= F_e(x_2,x_3) - M^{-1}F(x_1,x_3,\tilde{p}+\bar{p},t), \\
  \dot{\xi} &= C_t(r_3-x_3), \label{eq:errorb}\\
  \dot{\tilde{p}} &= C_h(\tilde{p}_\infty - \tilde{p}) \label{eq:errorc}.
\end{align}
Such error system can be seen in Fig. \ref{fig:Fulldiagram} and can be explained as the interconnection of two systems as in Fig. \ref{fig:Feedback}: system $H_1$ is defined as \eqref{eq:errorc} with $u_1=0_{2 n}$, $e_1=[-x_1^T,-x_3^T]^T$, and $y_1=[0_{1 \times n},(b(x_1,x_3,t) C_p \tilde{p})^T]^T$, and system $H_2$ is defined as \eqref{eq:errora}-\eqref{eq:errorb} with $u_2=0_{2 n}$, $e_2=[0_{1 \times n},(b(x_1,x_3,t) C_p \tilde{p})^T]^T$, and $y_2=[x_1^T,x_3^T]^T$.

\begin{thm}
Every bounded solution $(x(t),\xi(t),p(t))$ of the closed-loop system \eqref{eq:shift}, \eqref{eq:actdyn} and \eqref{eq:integral} approaches to the domain $E_p= \{\ (x,\xi,p) \in \Re^{3n}\times \Re^q \times \Re^q \mid C_t(r_3-x_3) =0_q, p = \bar{p} \}\ $ as $t \rightarrow \infty$ if the control input $p_\infty(x_1,x_3,\xi)$ is defined as 
\begin{equation}
\begin{split}
  p_\infty = \bar{p}_\infty + \tilde{p}_\infty, \quad
  \tilde{p}_\infty = -\mu_{min} C_p^T \abs{x_3},
\end{split}
\label{eq:pinf}
\end{equation}
with the nominal $\bar{p}_\infty$ defined as \eqref{eq:pinfnom}, fulfilling the conditions \eqref{eq:cond1}, $\lambda_\xi>0$ and matrix $C_t$ for the integral action \eqref{eq:integral} defined as in \eqref{eq:Ct}.
\label{th3}
\end{thm}

\begin{pf}
System \eqref{eq:errora}-\eqref{eq:errorb} is passive with the nominal control $\bar{p}$ as shown in the previous regulation analysis. Therefore, the condition $e_2^T y_2 \geq \dot{V} + \dot{V}_\xi + x_{3e}^T \bar{H} x_{3e}$ is fulfilled. Thus, the passivity property must be studied now in system \eqref{eq:errorc}.

Defining the positive definite storage function $V_p= \frac{1}{2} \tilde{p}^T C_h^{-1} \tilde{p}$ for the system \eqref{eq:errorc}, its derivative reads as
\begin{equation*}
\begin{split}
  \dot{V}_p &= \dot{\tilde{p}}^T C_h^{-1} \tilde{p}
  = (\tilde{p}_\infty - \tilde{p})^T C_h^T C_h^{-1} \tilde{p} \\
  &= -\mu_{min} x_3^T \sign{(x_3)} C_p \tilde{p} - \tilde{p}^T \tilde{p}
  \leq e_1^T y_1 - \tilde{p}^T \tilde{p},
\end{split}
\end{equation*}
resulting to be strictly passive and, consequently, the feedback connection between the two systems will be passive.

Finally, to get the domain in which the trajectories will converge, we use Theorem \ref{th:passive} to obtain $E_p= \{\ (x,\xi,p) \in \Re^{3n}\times \Re^q \times \Re^q \mid x_{3e}^T {H}^{or} x_{3e} +\tilde{p}^T \tilde{p} =0_{q}\}\ $, where the only possibility for the latter domain to be valid is if it takes the form of the one given in Theorem \ref{th3}. \hfill $\blacksquare$
\end{pf}

\section{Earthquake Control}
\label{sec:SimExp}

\begin{figure}[ht!]
  \centering 
  \includegraphics[width=7cm,height=5.5cm]{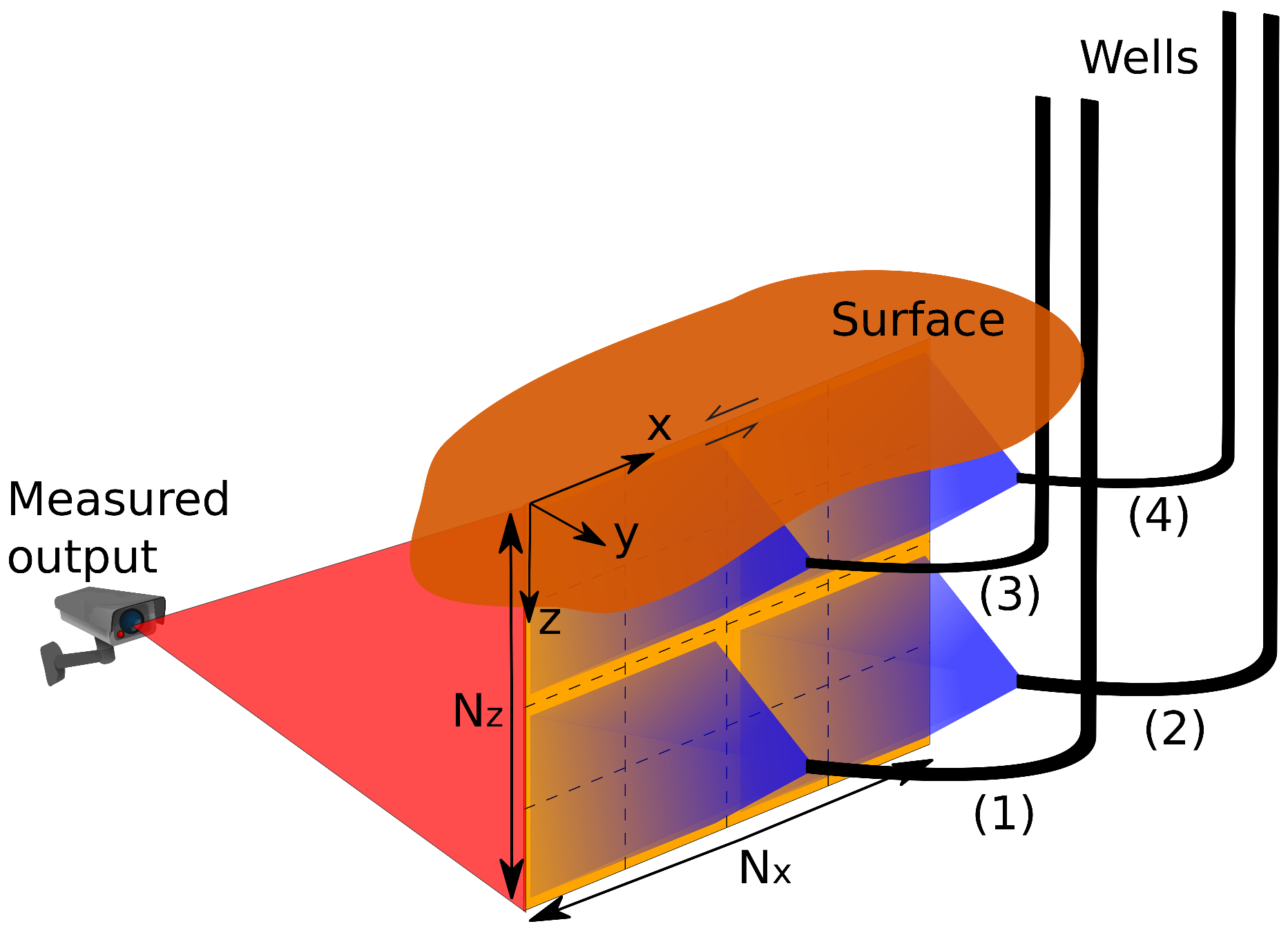}
  \caption{Illustration of a mature seismic fault discretized in $N_x \times N_z$ elements with four injection wells (inputs).}
  \label{fig:fault1}
\end{figure}

Consider a seismic fault as shown in Fig. \ref{fig:fault1}. In this academic example, the fault is just beneath the surface and its dimensions are $A=3 \times 3$ [km$^2$] (x- and z- directions, respectively). The effective normal stress $\sigma_n^{\prime}$ acting on the fault interface is assumed to vary linearly due to the lateral earth pressure. We assume also that the fault is adequately oriented in the tectonic stress regime for slip to occur. In this numerical application, the fault area is discretized into $n=N_x \times N_z = 10\times10$ elements.

The above physical system can be described mathematically using eqs. \eqref{eq:Fr}, \eqref{eq:shift}, \eqref{eq:Fr2}, and \eqref{eq:Coulomb2}, where $x_1$ represents the slip, $x_2$ the displacement and $x_3$ the slip-rate (velocity). Several methods in the literature can be used in order to discretize the differential operator representing the underlying continuum elastodynamic problem of seismic slip (\textit{e.g.}, Finite Element Method, Finite Differences, Boundary Element Method, spectral methods, model reduction methods, among others \cite{b:Barbot-2019}, \cite{b:Boyd-2000}, \cite{b:Erickson-etal-2020} and \cite{b:Larochelle-etal-2021}). In most cases, the resulting discretized equations will finally take the form of \eqref{eq:shift} and, consequently, the control theory presented in this work can be applied.

The actuator dynamics \eqref{eq:actdyn} is also considered, where the control input $p_\infty$ represents the pressure at the peak of four wells injecting fluid to the fault ($q=4$), as shown in Fig. \ref{fig:fault1}. The form of eq. \eqref{eq:actdyn} corresponds to a finite difference approximation of the diffusion equation, a Partial Differential Equation (PDE). Extension to PDE control could also be explored \cite{b:Gutierrez-Orlov-Plestan-Stefanou-CDC2022,b:Gutierrez-Orlov-Plestan-Stefanou-ECC2022,b:Krstic-Smyshlyaev-2008}, but this is out of the scope of the current work. The theorems developed in the previous section can be applied as the diffusion equation remains passive. Then, through the diffusion process according to equation \eqref{eq:actdyn}, the pressure $p$ affects the fault friction by modifying the effective normal stress $\sigma_n^{\prime}$ according to Terzaghi's principle of effective stress \cite{b:Terzaghi-1943}. The control configuration of the wells on the fault can be seen in Fig. \ref{fig:fault1}, where their influence is defined by the definition of matrix $C_p$ in the friction term \eqref{eq:Coulomb2}.

Furthermore, an even more realistic scenario will be studied where the full state $x(t)$ is not available, but only a measured output of the system \eqref{eq:shift} as $y_m = C_m x_3$, where $y_m \in \Re$ and $C_m \in \Re^{1 \times n}$. This single output represents an average velocity over the points of the fault. Therefore, the designed pressure at the fault $p=p(x)$ and, consequently, the designed pressure at the wells, $p_\infty=p_\infty(x)$, have to be now a feedback of the estimated states, \textit{i.e.} $\hat{p}=p(\hat{x})$ and $\hat{p}_\infty=p_\infty(\hat{x})$, respectively. The design of a high-gain observer for this purpose is shown in Appendix \ref{app:high-gain}.

\begin{figure*}[ht!]
  \centering 
  \includegraphics[width=0.33\textwidth,height=0.4\textheight,keepaspectratio]{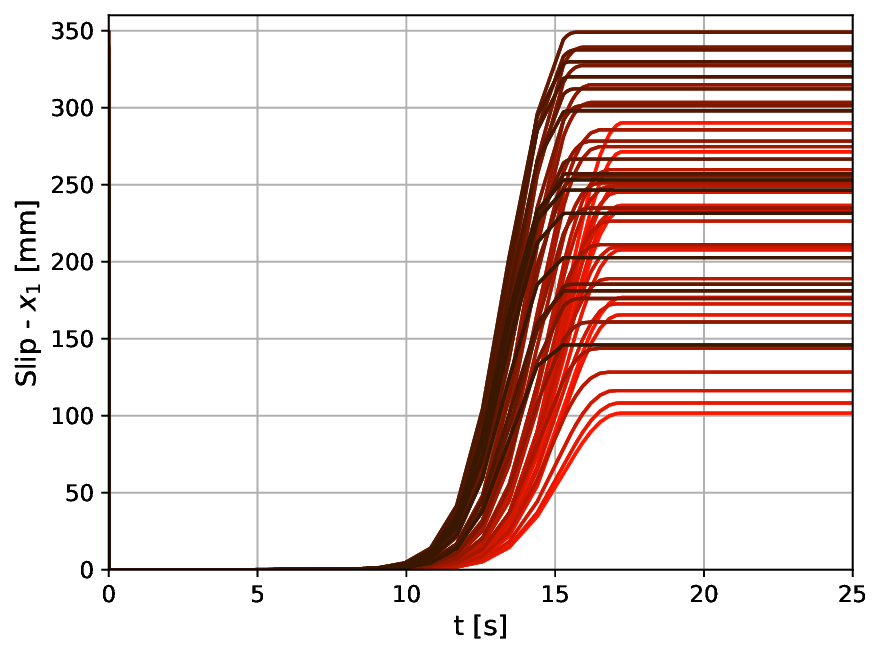}
  \includegraphics[width=0.33\textwidth,height=0.4\textheight,keepaspectratio]{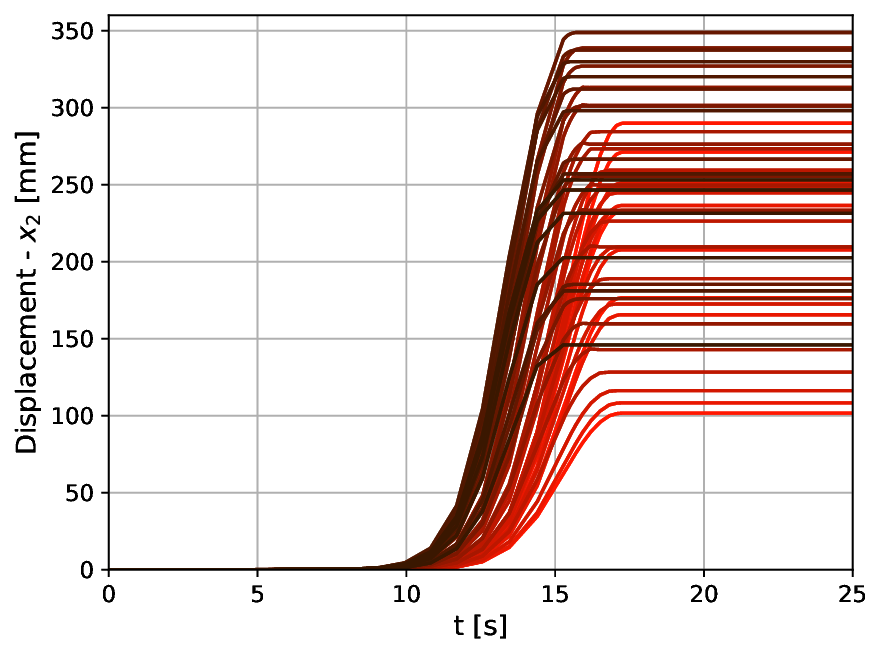}
  \includegraphics[width=0.33\textwidth,height=0.4\textheight,keepaspectratio]{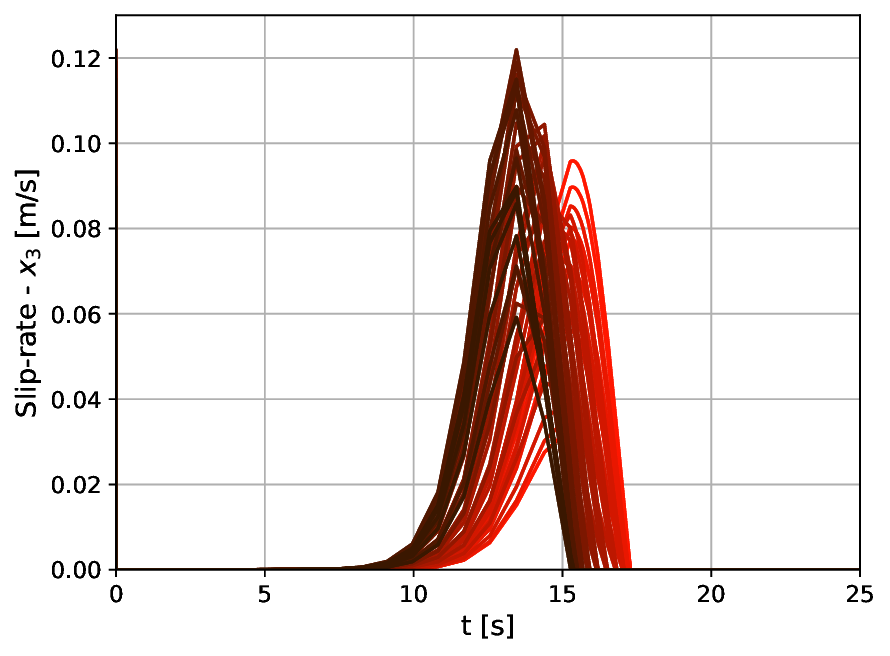}
  \caption{Earthquake-like behaviour showing fast slip dynamics (instability). Each curve represents an element of the discretized seismic fault and its color varies linearly with depth from red (depth $z=0$ [km]) to black (depth $z=3$ [km]).}
  \label{fig:x_no}
\end{figure*}

Without a control input, system \eqref{eq:shift} is unstable, resulting in an earthquake as shown in Fig. \ref{fig:x_no} (notice the time scale in seconds). It is worth mentioning that very few works are devoted to the control of such systems. In particular, an LQR control was designed to stabilize and perform tracking of an earthquake modelled by a MIMO system in \cite{b:Stefanou2019}, whereas a double-scale asymptotic approach was employed to design a transfer function-based control in \cite{b:Stefanou2020}. These first applications of control theory to this problem have shown that earthquakes could be controlled, at least from a mathematical point of view, but they have not accounted for underactuation, the discontinuous nature of friction and diffusion. Therefore, the presented theoretical development a more realistic treatment of the problem.

The objective in the sequel is to implement the designed control law \eqref{eq:pinfnom} and \eqref{eq:pinf} with the integral dynamics \eqref{eq:integral} and \eqref{eq:Ct}, to drive the system states to the domain $E_p= \{\ (x,\xi,p) \in \Re^{3n}\times \Re^q \times \Re^q \mid C_t(r_3-x_3) =0_q, p = \bar{p} \}\ $ as $t \rightarrow \infty$. If one chooses a small velocity reference $r_3$, this will result in a slow-aseismic response of the system.

The desired reference $r_3$ is a smooth function reading as
\begin{equation}
    r_3=\dot{r}(t)I_{n}, \quad r(t) = d_{max}s^3(10-15s+6s^2),
  \label{eq:ref}
\end{equation}
where $s = t/t_{op}$, $d_{max}$ is the target displacement and $t_{op}$ is the operational time of the tracking strategy. The constant $d_{max}$ is the distance the fault slides dynamically in order to reach its sequent stable equilibrium point. For this case, we selected $d_{max}=500$ [mm] (approximately two times equal to the seismic slip developed when the system is not controlled) and $t_{op}=360$ [days]. The desired total time is considerably larger than the fast slip in the earthquake behaviour ($\simeq 15$ [s]) in order to slowly release and dissipate the seismic energy. Shorter $t_{op}$ can be chosen as well (\textit{e.g.}, of the order of hours) but in this case, the pressure at the tips of the wells, $p_\infty$, would be very high due to slow dynamics of the diffusion process (see \eqref{eq:actdyn}). The characteristic time of the diffusion process (see \eqref{eq:actdyn}) depends on the hydraulic diffusivity parameter, which has been taken equal to $C_h=2.88\times 10^{-7}I$ (representing injection in a sandstone) and a distance of the injection point to the fault equal to $1.5$ [km].

\begin{rem}
The presented analysis for the regulation result in Section 5.2 fits only for constant references. Nevertheless, the resulting error could be improved by choosing references with low time derivatives, approximating its behaviour to constant references, like \eqref{eq:ref}. One can improve this result by adding more (passive) integrator terms to cover a wider range of references $r_3(t)$, as stated in the internal model principle (\textit{e.g.}, \cite{b:Gutierrez-Tzortzopoulos-Stefanou-Plestan-2022}).
\end{rem}

In this numerical example, we consider the friction coefficient $\mu(x_1+\delta_0,\abs{x_3},t)$ in \eqref{eq:Coulomb2} of the form $\mu_i(x_{1_i})=\mu_{res}-\Delta \mu \cdot e^{-\nicefrac{x_{1_i}}{d_c}}$, with $\Delta \mu<0$. Such function is defined as a slip-weakening friction law \cite{b:Kanamori-Brodsky-2004} and it evolves from an initial value $\mu_{max}$ (static friction coefficient), to a residual one $\mu_{res}$ (kinetic friction coefficient) in a characteristic slip $d_c$. Its values were chosen as $\mu_{res}=0.5$ (Assumption \ref{A4}), $\Delta \mu =\mu_{res}-\mu_{max}= 0.1$ and $d_c=10$ [m]. Other friction laws could be used as well (see \cite{b:https://doi.org/10.1029/2021JB023410}).

\subsection{Numerical Results}

In order to illustrate the performance of the proposed passivity-based control strategy, simulations have been made based on the shifted system described by \eqref{eq:Fr}, \eqref{eq:shift}, \eqref{eq:Fr2}, \eqref{eq:Coulomb2}, and \eqref{eq:actdyn}. Such simulations were performed using the Differential Equations package of Julia \cite{b:rackauckas2017differentialequations} and an initial condition $x(0)=0_{3n}$. In particular the TRBDF2 algorithm was used with events for detecting the transition between stick to slip and satisfy \eqref{eq:Fr}.

The control \eqref{eq:pinfnom} and \eqref{eq:pinf} with the integral dynamics \eqref{eq:integral} and \eqref{eq:Ct} were implemented in the simulations with $\lambda_\delta=40$ [Pa/m], $\lambda_v=346.4$ [Pa $\cdot$ s/m] and $\lambda_\xi=5 \times 10^{3}$ [Pa/m]. These gains were designed to satisfy \eqref{eq:cond1} with $\mu_{min}=\nicefrac{A\mu_{res}}{2}$ (Assumption \ref{A4}) and $l_\delta=4\Delta \mu/ d_c$, $l_v=0$ (Assumption \ref{A5}) due to the previously presented definition and parameters of the friction coefficient $\mu(x_1+\delta_0,\abs{x_3},t)$. The control uses the estimated states from the observer \eqref{eq:observer} with $\epsilon=0.1$ and an initial condition $\hat{x}(0)=0_{3n}$. 

\begin{figure*}[ht!]
  \centering 
  \includegraphics[width=0.33\textwidth,height=0.4\textheight,keepaspectratio]{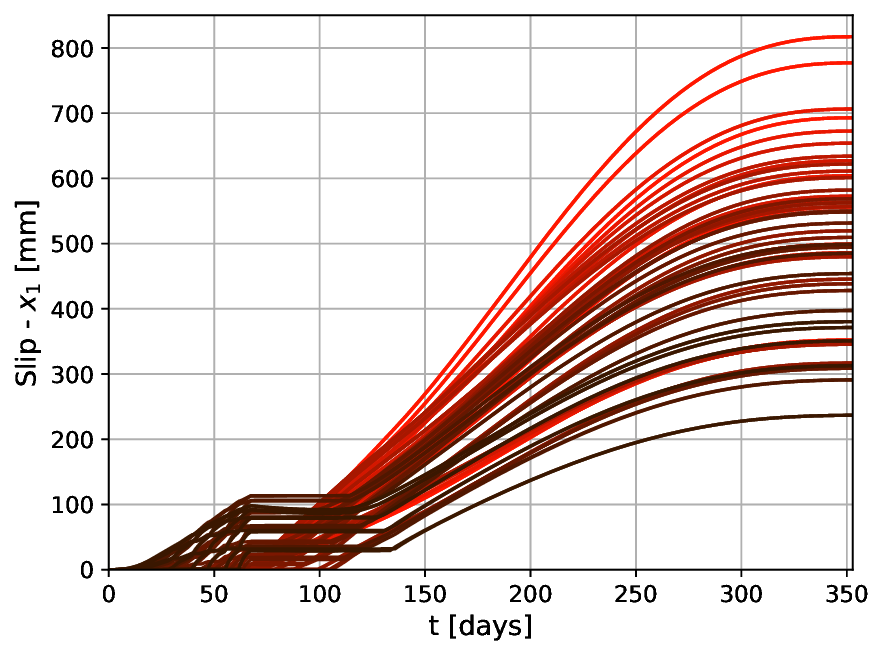}
  \includegraphics[width=0.33\textwidth,height=0.4\textheight,keepaspectratio]{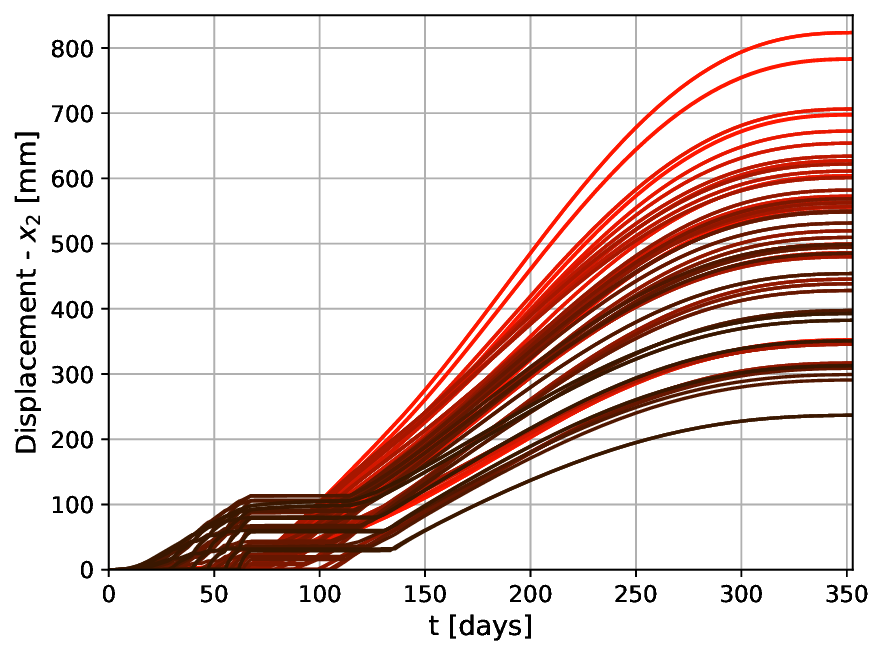}
  \includegraphics[width=0.33\textwidth,height=0.4\textheight,keepaspectratio]{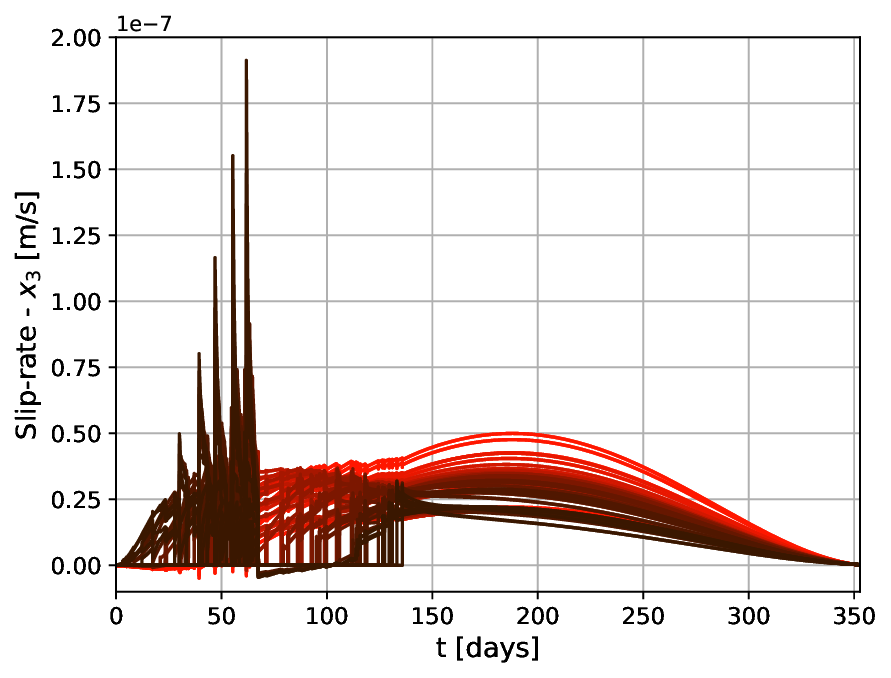}
  \caption{Controlled system: The system is tracked aseismically to a new (stable) equilibrium state. Note the difference on the time scale (days) with respect to the earthquake-like behaviour described in Fig. \ref{fig:x_no} (seconds). The same color code as Fig. \ref{fig:x_no} was used.}
  \label{fig:x}
\end{figure*}

\begin{figure*}[ht!]
  \centering 
  \includegraphics[width=0.33\textwidth,height=0.4\textheight,keepaspectratio]{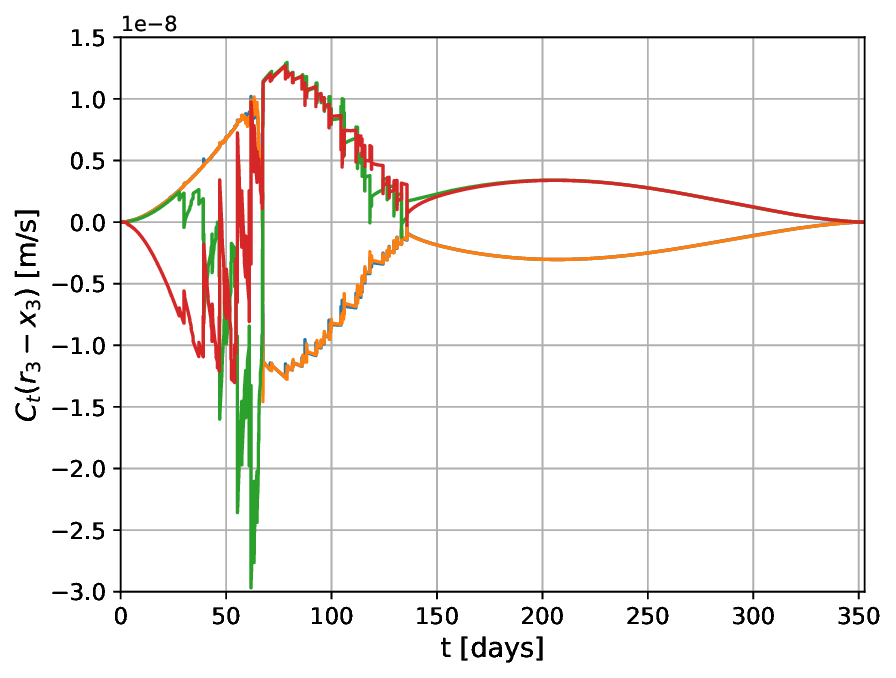}
  \includegraphics[width=0.33\textwidth,height=0.4\textheight,keepaspectratio]{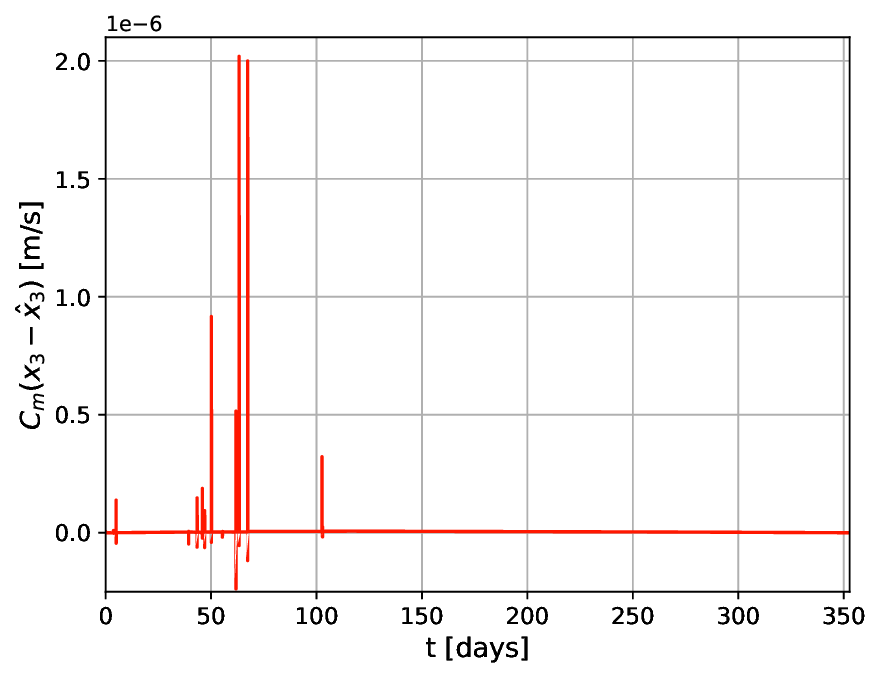}
  \includegraphics[width=0.33\textwidth,height=0.4\textheight,keepaspectratio]{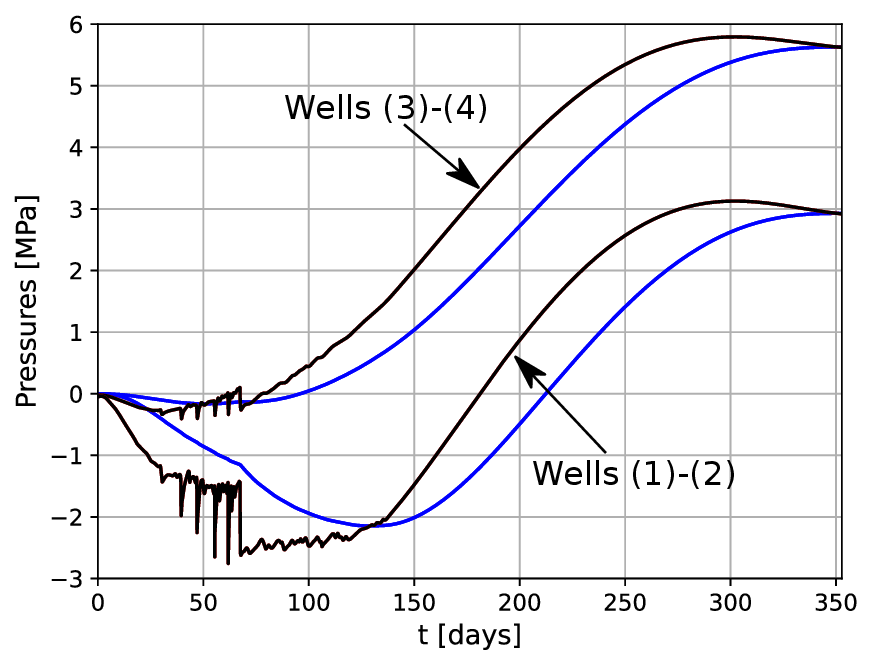}
  \caption{Integral (left) and observation (middle) errors. Control signals (right): Pressures developed on the seismic fault (blue), $p$, and control pressures applied on the wells (black), $p_\infty$. The delay is because of the slow dynamics of the actuator, due to diffusion.}
  \label{fig:p}
\end{figure*}

The results are presented in Figs. \ref{fig:x}-\ref{fig:p}. The states now follow successfully a slow reference, dissipating the stored energy aseismically (notice the time scale of days in Figs. \ref{fig:x}-\ref{fig:p} instead of seconds of the instability Fig. \ref{fig:x_no}). The discontinuous-like behaviour shown in the velocity $x_3$ is due to the stick-slip motion over the fault, resulting over the fact that Coulomb friction is a set-valued function (see \eqref{eq:Fr}, \eqref{eq:Fr2}, and \eqref{eq:Coulomb2}). Nevertheless, the designed control is able to drive the tracking error $C_t(r_3-x_3)$ close to zero, using the estimated states from the high-gain observer (errors shown in Fig. \ref{fig:p} left and middle plots). Finally, the control signal from the wells $p_\infty$ and the pressure $p$ applied to the fault are depicted in Fig. \ref{fig:p}, which show reasonable amplitudes to be used in real actuators.

\section{Conclusions}
\label{sec:Conclusions}

In this work, we extend the classic theorem for the negative feedback interconnection of passive systems to account for nonautonomous and set-valued (discontinuous) ODEs. This generalization is based on an invariance-like principle and it allows the synthesis of controllers for underactuated mechanical systems with Coulomb friction. Based on this generalization, stabilization of the states to a domain of zero velocities and tracking over constant references, while assuming actuation dynamics, are achieved. The designed control injects passivity to (unstable) frictional systems using less control inputs than degrees of freedom. It also need minimum information about the plant, \textit{i.e.}, the minimum bound of the friction coefficient, the belonging sector of the friction law and the coefficient of the actuator dynamics. This in contrast with the IDA-PBC where it is necessary to solve PDEs, or other existing more involved approaches. In order to test the derived control strategy, an earthquake prevention case study is considered. In particular, the unstable dynamic slip of a mature seismic fault is prevented by injecting fluid through four wells located far from the fault. Numerical simulations show the successful tracking of the system output over a reference, despite the presence of the slow dynamics due to diffusion process and uncertainties with respect to the Coulomb frictional rheology, the (visco-)elastodynamic properties of the system and diffusivity of the fluid pressure in the rock. 
The results were accomplished with minimum measurements and the control signals (pressures) were of acceptable amplitudes for the actuators (pumps). This results in a promising solution for earthquake prevention and control.

\begin{ack}                               
The authors would like to acknowledge the support of the European Research Council (ERC) under the European Union’s Horizon 2020 research and innovation program (Grant agreement no. 757848 CoQuake).
\end{ack}

\bibliographystyle{abbrv}        
\bibliography{Bibliografias}           

\begin{thebibliography}{10}

\bibitem{b:Adly-Le-2014}
S.~Adly and B.~K. Le.
\newblock {Stability and invariance results for a class of non-monotone
  set-valued Lur'e dynamical systems}.
\newblock {\em Applicable Analysis}, 93(5):1087–1105, 2014.

\bibitem{b:Armstromg-Dupont-Canudas-1994}
B.~Armstrong-H\'elovry, P.~Dupont, and C.~C.~D. Wit.
\newblock A survey of models, analysis tools and compensation methods for the
  control of machines with friction.
\newblock {\em Automatica}, 30(7):1083--1138, 1994.

\bibitem{b:AtassiKhalil_TAC1999}
A.~Atassi and H.~Khalil.
\newblock A separation principle for the stabilization of a class of nonlinear
  systems.
\newblock {\em {IEEE} Trans. Automat. Contr.}, 44(9):1672--1687, 1999.

\bibitem{b:10.1080/00207178708933801}
E.~Bailey and A.~Arapostathis.
\newblock Simple sliding mode control scheme applied to robot manipulators.
\newblock {\em International Journal of Control}, 45(4):1197--1209, 1987.

\bibitem{b:Barbot-2019}
S.~D. Barbot.
\newblock Slow-slip, slow earthquakes, period-two cycles, full and partial
  ruptures, and deterministic chaos in a single asperity fault.
\newblock {\em Tectonophysics}, 768:228171, 2019.

\bibitem{b:Boyd-2000}
J.~P. Boyd.
\newblock {\em Chebyshev and Fourier Spectral Methods: Second edition}.
\newblock Dover Publications, 2000.

\bibitem{b:Brogliato-2004}
B.~Brogliato.
\newblock {Absolute stability and the Lagrange–Dirichlet theorem with
  monotone multivalued mappings}.
\newblock {\em Systems \& Control Letters}, 51:343--353, 2004.

\bibitem{b:Brogliato-2022}
B.~Brogliato.
\newblock {Dissipative Dynamical Systems With Set-Valued Feedback Loops}.
\newblock {\em IEEE Control Systems Magazine}, 42(3):93--114, 2022.

\bibitem{b:10.1137/18M1234795}
B.~Brogliato and A.~Tanwani.
\newblock Dynamical systems coupled with monotone set-valued operators:
  Formalisms, applications, well-posedness, and stability.
\newblock {\em SIAM Review}, 62(1):3--129, 2020.

\bibitem{b:100932}
C.~Byrnes, A.~Isidori, and J.~Willems.
\newblock Passivity, feedback equivalence, and the global stabilization of
  minimum phase nonlinear systems.
\newblock {\em IEEE Transactions on Automatic Control}, 36(11):1228--1240,
  1991.

\bibitem{b:9740597}
N.~Chopra, M.~Fujita, R.~Ortega, and M.~W. Spong.
\newblock Passivity-based control of robots: Theory and examples from the
  literature.
\newblock {\em IEEE Control Systems Magazine}, 42(2):63--73, 2022.

\bibitem{b:10.1177/1077546311408469}
C.~Cornejo and L.~Alvarez-Icaza.
\newblock Passivity based control of under-actuated mechanical systems with
  nonlinear dynamic friction.
\newblock {\em Journal of Vibration and Control}, 18(7):1025--1042, 2012.

\bibitem{b:Canudas-Kelly-2007}
C.~C. de~Wit and R.~Kelly.
\newblock Passivity analysis of a motion controller for robot manipulators with
  dynamic friction.
\newblock {\em {Asian Journal of Control, Asian Control Association (ACA) and
  Chinese Automatic Control Society (CACS)}}, 9(9):30--36, 2007.

\bibitem{b:Erickson-etal-2020}
B.~A. Erickson, J.~Jiang, M.~Barall, N.~Lapusta, E.~M. Dunham, R.~Harris, and
  M.~Wei.
\newblock {The community code verification exercise for Simulating Sequences of
  Earthquakes and Aseismic Slip (SEAS)}.
\newblock {\em Seismological Research Letters}, 91(2A):874--890, 2020.

\bibitem{b:filippov}
A.~Filippov.
\newblock {\em Differential Equations with Discontinuous Right-hand Sides}.
\newblock Kluwer Academic Publishers, Dordrecht, The Netherlands, 1988.

\bibitem{b:Finogenko-2016}
I.~A. Finogenko.
\newblock The invariance principle for nonautonomous differential equations
  with discontinuous right-hand side.
\newblock {\em Siberian Mathematical Journal}, 57(4):715--725, 2016.

\bibitem{b:Franco-2021}
E.~Franco.
\newblock {IDA-PBC with adaptive friction compensation for underactuated
  mechanical systems}.
\newblock {\em International Journal of Control}, 94(4):860--870, 2021.

\bibitem{b:Gutierrez-Orlov-Plestan-Stefanou-CDC2022}
D.~Guti\'errez-Oribio, Y.~Orlov, I.~Stefanou, and F.~Plestan.
\newblock Robust motion planning for the heat equation using boundary control.
\newblock In {\em 61st {IEEE} Conference on Decision and Control}, Cancun,
  M\'exico, 2022.

\bibitem{b:Gutierrez-Orlov-Plestan-Stefanou-ECC2022}
D.~Guti\'errez-Oribio, Y.~Orlov, I.~Stefanou, and F.~Plestan.
\newblock Tracking for a wave equation using homogeneous boundary control.
\newblock In {\em 20th European Control Conference}, London, UK, 2022.

\bibitem{b:Gutierrez-Tzortzopoulos-Stefanou-Plestan-2022}
D.~Guti\'errez-Oribio, G.~Tzortzopoulos, I.~Stefanou, and F.~Plestan.
\newblock {Earthquake Control: An Emerging Application for Robust Control.
  Theory and Experimental Tests}.
\newblock {\em arXiv:2203.00296}, 2022.

\bibitem{b:1101352}
D.~Hill and P.~Moylan.
\newblock The stability of nonlinear dissipative systems.
\newblock {\em IEEE Transactions on Automatic Control}, 21(5):708--711, 1976.

\bibitem{b:HILL1977377}
D.~Hill and P.~Moylan.
\newblock Stability results for nonlinear feedback systems.
\newblock {\em Automatica}, 13(4):377--382, 1977.

\bibitem{b:Kamalapurkar-Rosenfeld-Parikh-Teel-Dixon-2019}
R.~Kamalapurkar, J.~A. Rosenfeld, A.~Parikh, A.~R. Teel, and W.~E. Dixon.
\newblock Invariance-like results for nonautonomous switched systems.
\newblock {\em IEEE Transactions on Automatic Control}, 64(2):614--627, 2019.

\bibitem{b:Kanamori-Brodsky-2004}
H.~Kanamori and E.~E. Brodsky.
\newblock The physics of earthquakes.
\newblock {\em Reports on Progress in Physics}, 67(8):1429--1496, 2004.

\bibitem{b:10.1080/00207178908953515}
R.~Kelly, R.~Carelli, and R.~Ortega.
\newblock Adaptive motion control design of robot manipulators: an input-output
  approach.
\newblock {\em International Journal of Control}, 50(6):2563--2581, 1989.

\bibitem{b:Khalil2002}
H.~Khalil.
\newblock {\em Nonlinear Systems}.
\newblock Prentice Hall, New Jersey, U.S.A., 2002.

\bibitem{b:Krstic-Smyshlyaev-2008}
M.~Krstic and A.~Smyshlyaev.
\newblock {\em Boundary Control of PDEs}.
\newblock SIAM Advances in Design and Control, 2008.

\bibitem{b:Larochelle-etal-2021}
S.~Larochelle, N.~Lapusta, J.~P. Ampuero, and F.~Cappa.
\newblock Constraining fault friction and stability with fluid-injection field
  experiments.
\newblock {\em Geophysical Research Letters}, 48(10):874--890, 2021.

\bibitem{b:Olsson-Astrom-Canudas-Gafvert-Lischinsky-1998}
H.~Olsson, K.~J. Astr\"om, C.~C. de~Wit, M.~G\"afvert, and P.~Lischinsky.
\newblock Friction models and friction compensation.
\newblock {\em European Journal of Control}, 4(3):176--195, 1998.

\bibitem{b:Orlov-2009}
Y.~V. Orlov.
\newblock {\em Discontinuous Systems: Lyapunov Analysis and Robust Synthesis
  under Uncertainty Conditions}.
\newblock Springer-Verlag, London, UK, 2009.

\bibitem{b:Ortega_TAC2002}
R.~Ortega, M.~Spong, F.~Gomez-Estern, and G.~Blankenstein.
\newblock Stabilization of a class of underactuated mechanical systems via
  interconnection and damping assignment.
\newblock {\em {IEEE} Transactions on Automatic Control}, 47(8):1218--1233,
  2002.

\bibitem{b:ORTEGA1989877}
R.~Ortega and M.~W. Spong.
\newblock Adaptive motion control of rigid robots: A tutorial.
\newblock {\em Automatica}, 25(6):877--888, 1989.

\bibitem{b:4048815}
B.~E. Paden and S.~S. Sastry.
\newblock {A calculus for computing Filippov's differential inclusion with
  application to the variable structure control of robot manipulators}.
\newblock In {\em 25th IEEE Conference on Decision and Control}, pages
  578--582, 1986.

\bibitem{b:Pennestri-Rossi-Salvini-Valentini-2016}
E.~Pennestr\'i, V.~Rossi, P.~Salvini, and P.~P. Valentini.
\newblock Review and comparison of dry friction force models.
\newblock {\em Nonlinear Dyn}, 83:1785--1801, 2016.

\bibitem{b:rackauckas2017differentialequations}
C.~Rackauckas and Q.~Nie.
\newblock Differentialequations.jl--a performant and feature-rich ecosystem for
  solving differential equations in julia.
\newblock {\em Journal of Open Research Software}, 5(1), 2017.

\bibitem{b:doi.org/10.1002/asjc.2718}
M.~Ruderman.
\newblock Stick-slip and convergence of feedback-controlled systems with
  coulomb friction.
\newblock {\em Asian Journal of Control}, pages 1--11, 2021.

\bibitem{b:bams/1183523748}
H.~J. Ryser.
\newblock {Matrices of zeros and ones}.
\newblock {\em Bulletin of the American Mathematical Society}, 66(6):442--464,
  1960.

\bibitem{b:doi.org/10.1002/rnc.1622}
J.~Sandoval, R.~Kelly, and V.~Santibáñez.
\newblock Interconnection and damping assignment passivity-based control of a
  class of underactuated mechanical systems with dynamic friction.
\newblock {\em International Journal of Robust and Nonlinear Control},
  21(7):738--751, 2011.

\bibitem{b:Spong-Vidyasagar}
M.~Spong and M.~Vidyasagar.
\newblock {\em Robot dynamics and control}.
\newblock Wiley, New York, USA, 1989.

\bibitem{b:Stefanou2019}
I.~Stefanou.
\newblock Controlling anthropogenic and natural seismicity: Insights from
  active stabilization of the spring-slider model.
\newblock {\em Journal of Geophysical Research: Solid Earth},
  124(8):8786--8802, 2019.

\bibitem{b:Stefanou2020}
I.~Stefanou.
\newblock Control instabilities and incite slow-slip in generalized
  burridge-knopoff models.
\newblock {\em arXiv:2008.03755}, 2020.

\bibitem{b:https://doi.org/10.1029/2021JB023410}
I.~Stefanou and G.~Tzortzopoulos.
\newblock Preventing instabilities and inducing controlled, slow-slip in
  frictionally unstable systems.
\newblock {\em Journal of Geophysical Research: Solid Earth},
  127(7):e2021JB023410, 2022.

\bibitem{b:Terzaghi-1943}
K.~Terzaghi.
\newblock {\em Theoretical Soil Mechanics}.
\newblock John Wiley \& Sons, Inc., 1943.

\bibitem{b:Willems-1972}
J.~C. Willems.
\newblock {Dissipative dynamical systems part I: General theory}.
\newblock {\em Arch. Rational Mech. Anal}, 45:321--351, 1972.

\end{thebibliography}

\appendix

\section{High-gain Observer Design}
\label{app:high-gain}

A high-gain observer design will be derived (see \cite[Chapter 14]{b:Khalil2002},\cite{b:AtassiKhalil_TAC1999}) to obtain the estimates states $\hat{x}$ of system \eqref{eq:shift} as
\begin{equation}
\begin{split}
  \dot{\hat{x}}_1 &= \abs{\hat{x}_3}, \\
  \dot{\hat{x}}_2 &= \hat{x}_3 + \lambda_1 L_1 (y_m-C_m \hat{x}_3), \\
  \dot{\hat{x}}_3 &= \hat{F}_e(\hat{x}_2,\hat{x}_3)- M_0^{-1}\hat{F}_r(\hat{x}_1,\hat{x}_2,\hat{x}_3,\hat{p},t)\\
  &\quad + \lambda_2 L_2 (y_m-C_m \hat{x}_3),
\end{split}
\label{eq:observer}
\end{equation}
where $\lambda_1,\lambda_2 \in \Re$, $L_1,L_2 \in \Re^{n \times m}$ are gains to be designed, $\hat{F}_e(\hat{x}_2,\hat{x}_3)=-K_0\hat{x}_2-H_0 \hat{x}_3$ and $K_0,H_0,M_0$, $\hat{F}_r(\hat{x}_1,\hat{x}_2,\hat{x}_3,\hat{p},t)$ are the the nominal matrices of $K,H,M$ and the nominal function of $F_r(x_1,x_2,x_3,\hat{p},t)$, respectively. 

Based on \cite{b:AtassiKhalil_TAC1999} and \cite[Chapter 14]{b:Khalil2002}, the estimation error $\tilde{x}=x-\hat{x}$ can be proved to be ISS with respect to the uncertain term 
$\delta(x,\hat{x})=-(K-K_0)x_2-(H-H_0)x_3-M^{-1}F_r(x_1,x_2,x_3,\hat{p},t)+M_0^{-1}\hat{F}_r(\hat{x}_1,\hat{x}_2,\hat{x}_3,\hat{p},t)$, if the observer gains are designed as $\lambda_1 = \frac{1}{\epsilon}$, $\lambda_2 = \frac{1}{\epsilon^2}$, with $\epsilon \approx 0$ and $L$ chosen to make the matrix $\tilde{A}=\left[\begin{array}{cc}
  0_{n\times n} & I_{n \times n}-L_1 C_m \\ 
  -K_0 & -H_0- L_2 C_m
  \end{array}  \right]$ Hurwitz. 

Is it worth noticing that the separation principle for nonlinear systems (\textit{e.g.}, \cite{b:AtassiKhalil_TAC1999}) consider systems with sufficiently smooth right-hand sides. Therefore, the analysis of the full closed loop-system (plant, control and high-gain observer) with discontinuous RHS presented in this paper, remains as future work.


\end{document}